\documentclass[10pt,aps,prd,twocolumn,showpacs,superscriptaddress,eqsecnum,longbibliography,nofootinbib]{revtex4-1}
\usepackage{mathrsfs,amsmath,amsthm,latexsym,amssymb,amsfonts,epsfig,txfonts,cancel,enumerate,graphicx,subfigure}
\newcommand{\makeSymbol}[1]{\mathord{\vcenter{\hbox{#1}}}}
\usepackage{xcolor}
\definecolor{navy}{RGB}{0,0,150}
\usepackage[colorlinks, linkcolor=navy, citecolor=navy, urlcolor=navy, plainpages=false, pdfstartview=FitH]{hyperref}
\usepackage{appendix}
\allowdisplaybreaks
\newcommand{\GZU}{School of Physics, Guizhou University, Guiyang 550025, China}
\newcommand{\UOW}{Faculty of Physics, University of Warsaw, Pasteura 5, 02-093 Warsaw, Poland}
\newcommand{\BNU}{Department of Physics, Beijing Normal University, Beijing 100875, China}

\begin{document}

\title{Relating spin-foam to canonical loop quantum gravity by graphical calculus}

\author{Jinsong Yang}
\email{jsyang@gzu.edu.cn}
\affiliation{\GZU}

\author{Cong Zhang}
\email{Cong.Zhang@fuw.edu.pl}
\affiliation{\UOW}

\author{Yongge Ma}
\thanks{Corresponding author}
\email{mayg@bnu.edu.cn}
\affiliation{\BNU}


\begin{abstract} 

The graphical calculus method is generalized to study the relation between covariant and canonical dynamics of loop quantum gravity. On one hand, a graphical derivation of the partition function of the generalized Euclidean Engle-Pereira-Rovelli-Livine (EPRL) spin-foam model is presented. On the other hand, the action of a Euclidean Hamiltonian constraint operator on certain spin network states is calculated by graphical method. It turns out that the EPRL model can provide a rigging map such that the Hamiltonian constraint operator is weakly satisfied on certain physical states for the Immirzi parameter $\beta=1$. In this sense, the quantum dynamics between the covariant and canonical formulations are consistent to each other.

\end{abstract}


\maketitle

\section{Introduction}

Loop quantum gravity (LQG) provides a nonperturbative and background-independent approach to the quantization of general relativity (GR). In the past thirty years, remarkable achievements have been made in the field of LQG  (see \cite{Rovelli:2004tv,Thiemann:2007pyv,Rovelli:2014ssa,Ashtekar:2017awx} for books, and \cite{Thiemann:2002nj,Ashtekar:2004eh,Han:2005km,Giesel:2012ws,Baez:1999sr,Rovelli:2011eq,Perez:2012wv} for review articles). Both the canonical and the covariant (path integral) formulations of LQG have been developed.

Canonical LQG is based on the Hamiltonian formulation of GR in the Ashtekar-Barbero variables \cite{Ashtekar:1986yd,Ashtekar:1987gu,Barbero:1994ap}. The spacetime manifold has the structure $M\cong\varmathbb{R}\times \Sigma$ with $\Sigma$ being a 3-dimensional manifold of arbitrary topology. The canonical variables defined on $\Sigma$ are the $su(2)$-valued connection $A^i_a(x)$ and the densitized triad $\tilde{E}^a_i(x)$, where $i,j,k,\cdots=1,2,3$ are the $su(2)$ indices while $a,b,c$ are the spatial indices. The only nontrivial Poisson bracket between these variables reads
\begin{align}\label{eq}
\{A^i_a(x),\tilde{E}^b_j(y)\}=\kappa\,\beta\,\delta^b_a\delta^i_j\delta^3(x,y)\,,
\end{align}
where $\kappa\equiv8\pi G$ with $G$ being the Newtonian constant, and $\beta$ denotes the Immirzi parameter \cite{Barbero:1994ap,Immirzi:1996dr}. The elementary algebra, which can be directly promoted to that of the fundamental operators, consists of the holonomies $g_e(A)$ of $A_a^i$ along one-dimensional edges $e$ and fluxes $\tilde{E}_j(S)$ of $\tilde{E}^a_i$ through two-dimensional surfaces $S$. It turns out that there is a unique gauge and diffeomorphism invariant cyclic representation of the holonomy-flux $C^*$-algebra \cite{Lewandowski:2005jk}. The resulting representation space is the gauge and diffeomorphism invariant version of the kinematical Hilbert space ${\cal H}_{\rm kin}:=L^2(\bar{\cal A},{\rm d}\mu_o)$, where $\bar{\cal A}$ is the space of distributional connections, and ${\rm d}\mu_o$ is the Ashtekar-Lewandowski measure \cite{Ashtekar:1991kc,Ashtekar:1994mh}. The basis of ${\cal H}_{\rm kin}$ consists of the spin network states $T_{\gamma,\vec{j},\vec{i}}(A)$ defined on arbitrary finite graphs $\gamma$ in $\Sigma$ with a spin $j_e$ and an intertwiner $i_v$ coloring each edge $e$ and each vertex $v$ of $\gamma$. The classical spatial geometric functions, such as the length, area, and volume have been successfully quantized as the corresponding operators in ${\cal H}_{\rm kin}$, and they all have discrete spectra \cite{Rovelli:1994ge,Ashtekar:1996eg,Ashtekar:1997fb,Yang:2016kia,Thiemann:1996at,Ma:2010fy}. In the connection formulation, GR is cast into a constrained system with three first-class constraints, the Gaussian, diffeomorphism, and Hamiltonian constraints. The Gaussian and diffeomorphism constraints have been successfully implemented at quantum level. Thus the quantum dynamics is encoded in the Hamiltonian constraint. How to suitably quantize the Hamiltonian constraint and how to construct the physical Hilbert space are still under debate. Thus the quantum dynamics in canonical LQG remains obscure up to now. Nevertheless, some well-defined Hamiltonian constraint operators for pure gravity as well as gravity coupled to matter were constructed in different ways \cite{Thiemann:1996aw,Thiemann:1997rt,Yang:2015zda,Alesci:2015wla,Tomlin:2012qz,Varadarajan:2012re,Varadarajan:2018tei,Varadarajan:2019wpu}. Some properties of certain Hamiltonian operators were studied analytically as well as numerically \cite{Alesci:2011ia,Thiemann:2013lka,Zhang:2018wbc,Zhang:2019dgi}.

As a kind of path-integral formalism for GR, covariant LQG is well known as certain spin-foam model (SFM). A spin-foam is a dual 2-cell complex $\Delta^*$ with faces $f$ labeled by spins $j_f$ and edges $e$ labeled by intertwiners $i_e$. A slice of a spin-foam at ``fixed time'' gives a spin network state. Hence a spin-foam can be interpreted as an evolutional history of a spin network state, and can be understood as a formulation describing the quantum geometry of spacetime. A SFM is defined by assigning transition amplitudes $A_f$, $A_e$ and $A_v$ to the faces $f\in\Delta^*$, the edges $e\in\Delta^*$ and the vertices $v\in \Delta^*$, respectively. The key observation of current SFMs is that 4-dimensional GR can be written as a $BF$ theory with the so-called simplicity constraint forcing the $B$ field to be obtained from the tetrad field. Hence the strategy is first to derive the $BF$ partition function ${\cal Z}^{\rm BF}(\Delta^*)$ by discretizing the $BF$ action on $\Delta^*$ and its dual $\Delta$, and then to impose a quantum version of the discretized simplicity constraint on ${\cal Z}^{\rm BF}(\Delta^*)$, leading to the resulting partition function
\begin{align}
{\cal Z}^{\rm SFM}(\Delta^*)&=\sum_{j_f, i_e}\prod_fA_f\prod_eA_e\prod_vA_v.
\end{align}
Different implementing schemes of the simplicity constraint lead to different SFMs, for examples, the Barrett-Crane (BC) model \cite{Barrett:1997gw,Barrett:1999qw}, the Engle-Pereira-Rovelli-Livine (EPRL) model \cite{Engle:2007wy}, and the Freidel-Krasnov (FK) model \cite{Freidel:2007py}. The advantage of EPRL model and FK model is that they have correct classical limit to certain sense. The essential difference between the two models and BC model is that the simplicity constraint restrains $BF$ action to the Holst action in the formers, but to the Palatini action in the latter. The simplicity constraint was implemented differently in EPRL and FK models. In the former it was imposed at quantum level by the master-constraint criterion \cite{Engle:2007wy} or the Gupta-Bleuler criterion \cite{Ding:2009jq}, while in the latter it was imposed as a semiclassical condition on the coherent state basis proposed by Livine and Speziale \cite{Livine:2007vk}. The two models share the same vertex amplitude for $\beta\leqslant1$, but differ for $\beta>1$. Furthermore, the EPRL model was successfully generalized to the Kami{\'n}ski-Kisielowski-Lewandowski (KKL) model \cite{Kaminski:2009fm,Ding:2010fw}, which allows arbitrary boundary graphs. 

Whether the dynamics of covariant formulation is equivalent to that of canonical formulation is still an open issue in LQG up to now. Fortunately, the EPRL model generalized by KKL to arbitrary boundary graphs, supporting the quantum states of canonical LQG, has opened a door to set up the relation between the two formulations. Actually, it was shown in Ref. \cite{Alesci:2011ia} that the rigging map defined by the transition amplitude of EPRL SFM can give certain physical states of the quantum Euclidean Hamiltonian constraint $\hat{H}^{\rm E}_{\rm T}(N)$ of canonical LQG proposed by Thiemann in \cite{Thiemann:1996aw} for $\beta=1$ in the sense that the matrix elements of $\hat{H}^{\rm E}_{\rm T}(N)$ vanish. This implies a consistency of the quantum dynamics between covariant and canonical formulations for these states. The aim of this paper is to check whether such a consistency exists also between the EPRL SFM and the Hamiltonian constraint operator proposed in \cite{Yang:2015zda} for canonical LQG. We will consider only the Euclidean part of the Hamiltonian constraint in \cite{Yang:2015zda} and generalize the graphical calculus presented in \cite{Yang:2015wka}, which is based on the original Brink's graphical method, to deal with the explicit computations including the SFM. The graphical calculus has been systematically applied to canonical LQG with the virtues of concise and visual formulas, providing a powerful technique for simplifying the complicated calculations \cite{Brink:1968bk,Yang:2015wka,Yang:2016kia,Yang:2019xms}. Our results show that the rigging map of the Euclidean EPRL model with $\beta=1$ generalized to arbitrary boundary graphs does give certain physical states for the Euclidean Hamiltonian constraint operator defined in \cite{Yang:2015zda} with a special factor ordering, in the same sense as in Ref. \cite{Alesci:2011ia}.

The rest of this paper is organized as follows. In Sec. \ref{sec-II}, we give a detailed and concise derivation of the partition function of the generalized Euclidean EPRL model using graphical calculus, in parallel with the algebraic derivation in \cite{Ding:2010fw}. In Sec. \ref{sec-III}, we graphically calculate the action of the Euclidean Hamiltonian constraint operator defined in \cite{Yang:2015zda} with a special factor ordering on the spin network states $\psi_i$ with a 4-valent vertex $v$, and obtain its matrix elements. In Sec. \ref{sec-IV}, we show for $\beta=1$ that the rigging map of generalized Euclidean EPRL model can provide certain physical states of the Euclidean Hamiltonian constraint operator in \cite{Yang:2015zda} such that its matrix elements vanish. In this sense, the quantum dynamics between covariant and canonical LQG are again consistent for these states. Our results are summarized and discussed in Sec. \ref{sec-V}.

\section{The partition function in SFM}
\label{sec-II}
In this section, we will give a concise graphical derivation of the partition function for the Euclidean EPRL model gener alized by KKL. The starting point of SFMs is the fact that classical GR can be cast as a constrained $BF$ theory. Thus the strategy is to first derive the the partition function of $BF$ theory, and then to impose the simplicity constraint in a satisfactory manner. The partition function of $BF$ theory with gauge group $SO(4)$ or ${\rm Spin}(4)$ for the Euclidean case in 4-dimensions reads \cite{Rovelli:2004tv,Thiemann:2007pyv,Perez:2012wv}
\begin{align}\label{eq:Zbf}
 {\cal Z}^{\rm BF}(M)=\int {\rm d}A{\rm d}B \;e^{{\rm i}\int_{M}{\rm Tr}\left[B\wedge F(A)\right]}=\int {\rm d}A\,\prod_{x\in M}\delta\left[F(A)\right],
\end{align}
where $B$ is a $so(4)$-valued 2-form field on the spacetime manifold $M$, $F$ is the curvature of the $so(4)$ connection $A$ on $M$, the trace ${\rm Tr} $ is with respect to the Cartan-Killing metric on $so(4)$, and in the second step a formal integration over $B$ field leads to a Dirac delta function. To give a precise meaning to the formal expression \eqref{eq:Zbf}, one needs to discretize $M$ and employ its discrete structure. 

Suppose $M$ can be discretized by an arbitrary oriented 2-cell complex $\Delta$. We refer to Refs. \cite{Baez:1999sr,Rourke:1972bks,Kaminski:2009fm,Ding:2010fw} for the definition of  2-cell complex. The dual 2-cell complex $\Delta^*$ of $\Delta$ consists of 2-dimensional faces $f\in\Delta^*$, 1-dimensional edges $e\in\Delta^*$ and 0-dimensional vertices $v\in\Delta^*$. Denote by $\partial f$ the cyclically ordered set of edges bounding the face $f$ and the set of vertices bounding the boundary edges of $f$, by $\partial e$ the set of faces bounded by $e$, and by $\partial v$ the set of edges bounded by $v$ and the set of faces containing $v$ in their boundaries. For $\Delta^*$ with a boundary $\partial\Delta^*$, we mean that $\partial\Delta^*$ is a 1-cell complex, called the global boundary graph $\gamma\equiv\partial\Delta^*$, such that it is closed and does not contain any vertex of $\Delta^*$. An edge $e\in\partial\Delta^*$ is called an external edge (link) and it is contained in only one face. A vertex $v\in\partial \Delta^*$ is called an external vertex (node) and it is contained in exactly one internal edge of $\Delta^*$. Given an internal vertex $v\in\Delta^*$, the local boundary graph $\gamma_v$ of $v$ is the intersection between $\Delta^*$ and a small sphere surrounding $v$. The edges (links) of $\gamma_v$ are the intersections of $f\in\partial v$ with the sphere, denoted by $fv$, and the orientations of the edges are induced by those of $f$. The vertices (nodes) of $\gamma_v$  are the intersections of $e\in \partial v$ with the sphere. A spin-foam ${\cal F}$ is a triple $(\Delta^*, \vec{\rho},\vec{i})$ consisting of $\Delta^*$, a collection $\vec{\rho}$ of irreducible representations $\rho_f$ of ${\rm Spin}(4)$ assigned for each face $f\in\Delta^*$, and a collection $\vec{i}$ of intertwiners $i_e$ associated to each edge $e\in\Delta^*$. A SFM based on a spin-foam is defined by an assignment of the amplitudes $A_f$, $A_e$ and $A_v$ associated to the internal faces $f\in \Delta^*$, edges $e\in\Delta^*$ and vertices $v\in\Delta^*$. In the case that $\Delta^*$ has a boundary $\gamma\cup\gamma'$, a SFM contains also the boundary transition amplitude from the spin network state on $\gamma$ to the one on $\gamma'$.

\begin{figure*}[t]
\centering
\subfigure[~$\Delta^*$ without boundary]
{\label{Figa}\includegraphics[width=0.3\textwidth]{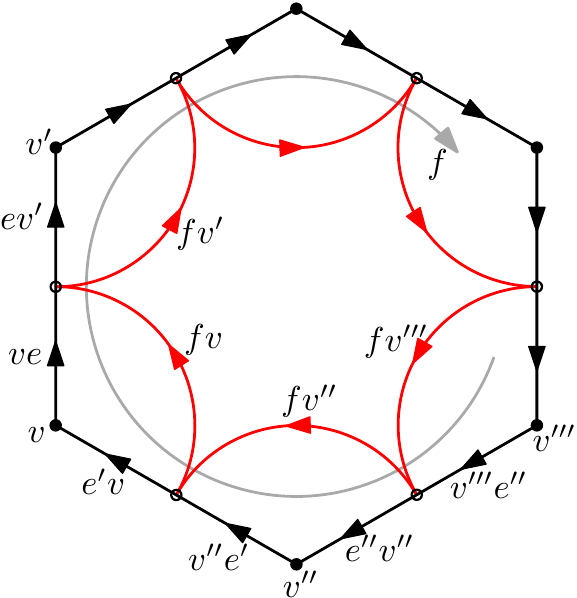}}
\hspace{2.5cm}
\subfigure[~$\Delta^*$ with a boundary ]
{\label{Figb}\includegraphics[width=0.3\textwidth]{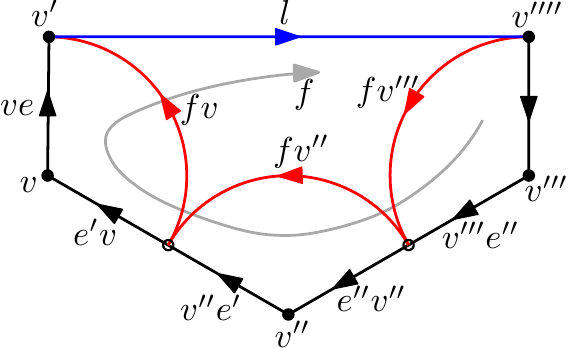}}
\caption{(a) A part of an oriented face $f$ of $\Delta^*$ for the case of $\Delta^*$ without boundary: The orientations of boundary edges $e\in\partial f$, represented by the arrows, are induced by that of $f$. The (internal) vertices (endpoints) $v$ of the (internal) edges $e$ are denoted by the solid points, while the midpoints of $e$ are represented by hollow circles, by which each edge $e$ is broken into two segments $ve$ and $ev'$ with orientations agreeing with that of $e$, and the oriented red curves lying in $f$ represent the edges $fv$ of the vertex-boundary graphs $\gamma_v$ based at $v$. (b) A part of $\Delta^*$ consisting of an oriented face $f$ bounded by a global boundary edge $l$ for the case of $\Delta^*$ with a boundary: The red curves denote again the edges $fv$ of the vertex boundary graph $\gamma_v$ lying in $f$ based at internal vertices $v$, and the blue curve represents the global boundary edge (link) $l$ with two vertices (nodes) $v'$ and $v''''$.}
\label{Fig}
\end{figure*}
The partition function $ {\cal Z}^{\rm BF}$ on $\Delta^*$ can be discretized. Given a $\Delta^*$ of $M$, approximating the curvatures $F(A)$ by holonomies $g_{\partial f}=\prod_{e\in\partial f}g_e$ around the loops $\partial f$ composed by the cyclically ordered sets of edges bounding the faces $f$, and replacing ${\rm d}A$ by the Haar measure ${\rm d}g_e$ on ${\rm Spin}(4)$, the discretized $BF$ partition function corresponding to Eq. \eqref{eq:Zbf} is defined by \cite{Ding:2010fw,Alesci:2011ia,Thiemann:2013lka}
\begin{align}\label{Z-BF-alg}
 &{\cal Z}^{\rm BF}(\Delta^*):=\int {\rm d}g_e\prod_{f\in\Delta^*}\delta\left(\prod_{e\in\partial f}g_eg_l\right)\notag\\
 &=\int {\rm d}g_{ve}\int{\rm d}g_{fv}\prod_{f\in\Delta^*}\delta\left(\prod_{v\in\partial f}g_{fv}g_l\right)
 \prod_{fv}\delta(g_{e'v}g_{ve}g_{fv}^{-1})\notag\\ 
 &=\int{\rm d}g^+_{fv}\int {\rm d}g^+_{ve}\prod_{f\in\Delta^*}\sum_{j^+_f}d_{j^+_f}{\rm Tr}_{j^+_f}\left(\prod_{v\in\partial f}g^+_{fv}g^+_l\right)\notag\\
 &\hspace{1.4cm}\times\prod_{fv}\sum_{j^+_{fv^{-1}}}d_{j^+_{fv^{-1}}}{\rm Tr}_{j^+_{fv^{-1}}}(g^+_{e'v}g^+_{ve}g^+_{fv^{-1}})\notag\\
 &\quad\times\int{\rm d}g^-_{fv}\int {\rm d}g^-_{ve}\prod_{f\in\Delta^*}\sum_{j^-_f}d_{j^-_f}{\rm Tr}_{j^-_f}\left(\prod_{v\in\partial f}g^-_{fv}g^-_l\right)\notag\\
 &\hspace{1.4cm}\times\prod_{fv}\sum_{j^-_{fv^{-1}}}d_{j^-_{fv^{-1}}}{\rm Tr}_{j^-_{fv^{-1}}}(g^-_{e'v}g^-_{ve}g^-_{fv^{-1}}),
\end{align}
where $g_e$ denotes holononmies along internal edges $e\in \partial f$ with orientations induced by $f$, $g_l$ denotes holonomies along boundary edges $l=f\cap\partial\Delta^*$ with  orientations induced by $f$ in the case that $\Delta^*$ has a boundary and we set $g_l=\mathbb{I}_{\rm Spin(4)}$ in the case that $\Delta^*$ has no boundary, in the second step, we split each internal edge $e$ bounded by $v$ and $v'$ into two segments $ve$ and $ev'$ with the same orientation as that of $e$, and regrouped the groups on segments of internal edges associated to $v$, and the auxiliary group elements $ g_{fv}$ are constrained by including the  additional delta functions, in the third step, we used the fact of ${\rm Spin}(4)\cong SU(2)\times SU(2)$ which allows us to expand the delta functions on $g\in {\rm Spin(4)}$ in terms of irreducible representations $(j^+,j^-)$ of $(g^+,g^-)\in SU(2)\times SU(2)$ as
\begin{align}\label{eq:delta-delta}
\delta(g)=\delta(g^+)\,\delta(g^-)=\sum_{j^+}d_{j^+}{\rm Tr}_{j^+}(g^+)\sum_{j^-}d_{j^-}{\rm Tr}_{j^-}(g^-),
\end{align}
with $d_j:=2j+1$ being the dimension of the representation space ${\cal H}_j$ of $SU(2)$ and the trace ${\rm Tr}_j$ being taken in the irreducible representation $\pi_j$ of $SU(2)$, and $fv^{-1}$ denotes the edge obtained from $fv$ by flipping its orientation. (See Fig. \ref{Fig} for a visual explanation of the notations).

To further derive the partition function ${\cal Z}^{\rm SFM}(\Delta^*)$ of a SFM for GR from ${\cal Z}^{\rm BF}(\Delta^*)$, one uses the following procedure \cite{Ding:2010fw}. First, integrating out the group elements $g_{ve}^\pm$ associated to the internal edges of $\Delta^*$ reduces the integrand in ${\cal Z}^{\rm BF}(\Delta^*)$ into a function $\prod_fA^{\rm BF}_f(\{g^+_{fv},g^-_{fv};g^+_l,g^-_l\})$. Second, the quantum simplicity constraint is imposed in a suitable way on the Hilbert spaces ${\cal H}_{\gamma_v}$ associated to all (local) vertex-boundary graphs $\gamma_v$. This will further restrict $A^{\rm BF}_f(\{g^+_{fv},g^-_{fv};g^+_l,g^-_l\})$ to $A^{\rm SFM}_f(\{g^+_{fv},g^-_{fv};g^+_l,g^-_l)$. Implementing the simplicity constraint in different manners results in different SFMs. Finally, for each internal vertex $v\in\Delta^*$, one performs the integration over the group elements $g_{fv}^\pm$ associated to $\gamma_v$. Then the resulting partition function ${\cal Z}^{\rm SFM}(\Delta^*)$ will be expressed as a sum over representations and intertwiners. Note that a derivation of the partition function by a procedure alternative to the above one was presented in \cite{Perez:2012wv}.

\subsection{The graphical calculus}
In this subsection, we briefly recall the elements of the Brink's graphical calculus applied in canonical LQG (see \cite{Brink:1968bk,Yang:2015wka} for details), and then extend it to compute the integral over the product of irreducible representations of $SU(2)$ in order to derive ${\cal Z}^{\rm SFM}(\Delta^*)$. We will focus on the graphical representations of the matrix elements of holonomies $g_e$ associated to edges $e$, the intertwiners $i_v$ associated to vertices $v$, and the graphical transformation rules.

The intertwiner is closely related to the $3j$-symbol, which is graphically represented by an oriented node with three black lines labeled by three angular momenta and a sign factor as
\begin{align}\label{3j-def-graph}
\begin{pmatrix}
j_1 & j_2 & j_3\\
m_1 & m_2 & m_3
\end{pmatrix}
&=\makeSymbol{
\includegraphics[width=2cm]{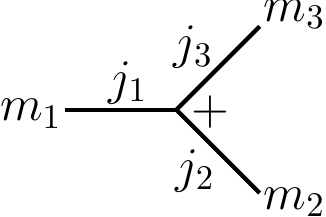}}=\makeSymbol{
\includegraphics[width=2cm]{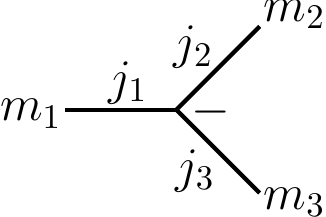}},
\end{align}
where the sign $-$ (or $+$) denotes the clockwise (or counterclockwise) orientation of the node with the cyclic order of the lines. A rotated diagram represents the same $3j$-symbol as the initial diagram, and the angles between two lines as well as the lengths of lines have no significance. A special $3j$-symbol with one zero-valued angular momentum is related to the ``metric" tensor $C^{(j)}_{mm'}$ graphically by
\begin{align}
\makeSymbol{
\includegraphics[width=1.9cm]{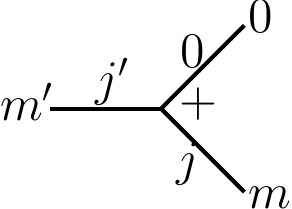}}=\frac{\delta_{j,j'}}{\sqrt{d_j}}\;\makeSymbol{
\includegraphics[width=1.9cm]{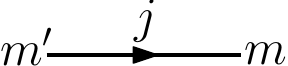}}\label{arrow-3j}\,,
\end{align}
where a black line with an arrow on it graphically represents the ``metric" tensor
\begin{align}\label{metric-graph}
C^{(j)}_{mm'}=(-1)^{j-m'}\delta_{m,-m'}=(-1)^{j+m}\delta_{m,-m'}=\makeSymbol{
\includegraphics[width=1.9cm]{figures/graphical-rules/wigner-3j-symbol-4}}\,.
\end{align}
The ``metric" tensor $C^{(j)}_{mm'}$ on ${\cal H}_j$ often occurs in the contraction of two $3j$-symbols with the same $j$ values. The inverse $C^{m'm}_{(j)}$ can be expressed by
\begin{align}
C^{m'm}_{(j)}=(-1)^{j-m'}\delta_{m,-m'}=(-1)^{j+m}\delta_{m,-m'}=\makeSymbol{
\includegraphics[width=1.9cm]{figures/graphical-rules/wigner-3j-symbol-4}}\,.
\end{align}
A black line denoted by $j$ without arrow on it represents the Kronecker delta in ${\cal H}_j$, i.e.,
\begin{align}\label{Kronecker-graph}
{{\delta^{(j)}}^m}_{m'}=\makeSymbol{
\includegraphics[width=1.9cm]{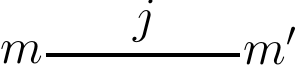}}\,.
\end{align}
The contraction of a $3j$-symbol with a ``metric'' represents the Clebsch-Gordan coefficient multiplied by a factor. In graphical representation, summation over the magnetic quantum numbers $m$ is represented by joining the free ends of the corresponding lines. Hence graphically the contraction of a $3j$-symbol with a ``metric'' is represented by a node with one arrow as
\begin{align}\label{CG-graph}
\makeSymbol{
\includegraphics[width=2.6cm]{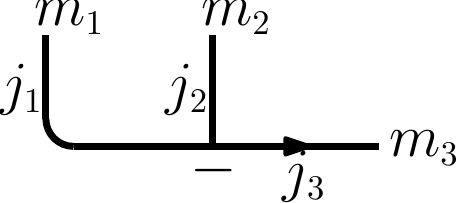}}&=\begin{pmatrix}
j_1 & j_2 & j_3\\
m_1 & m_2 & m'_3
\end{pmatrix}C^{m'_3m_3}_{(j_3)}\notag\\
&=\frac{(-1)^{j_1-j_2-j_3}}{\sqrt{d_{j_3}}}\langle j_3m_3| j_1m_1j_2m_2\rangle\,,
\end{align}
which is the building block in the construction of an intertwiner. Notice that the intertwiner is defined up to a factor with norm $1$. The normalized intertwiner $i_v$ associated to a vertex $v$, from which $n$ edges with $n$ spins $j_1,\cdots, j_n$ start, is defined by \cite{Yang:2015wka}
\begin{align}\label{intertw-true}
&{\left(i^{\,J;\,\vec{a}}_v\right)_{m_1m_2\cdots m_n}}^M\equiv{\left(i^{\,J;\,\vec{a}}_{j_1\cdots j_n}\right)_{\,m_1m_2\cdots m_n}}^M\notag\\
:=&(-1)^{j_1-\sum_{i=2}^nj_i-J}\langle JM;\vec{a}\;|\,j_1m_1j_2m_2\cdots j_nm_n\rangle\notag\\
=&(-1)^{j_1-\sum_{i=2}^nj_i-J}\sum_{k_2,\cdots,k_{n-1}}\langle a_2k_2|j_1m_1j_2m_2\rangle\langle a_3k_3|a_2k_2j_3m_3\rangle\notag\\
&\hspace{3cm}\times\cdots\times\langle JM|a_{n-1}k_{n-1}j_nm_n\rangle\notag\\
=&\prod_{i=2}^{n-1}\sqrt{d_{a_i}}\sqrt{d_J}\;
\makeSymbol{\includegraphics[width=5.7cm]{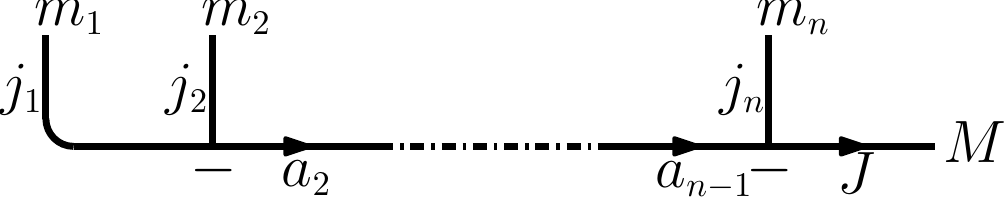}},
\end{align}
which describes the coupling of $n$ angular momenta $j_1,\cdots,j_n$ to a total angular momentum $J$ in the standard coupling scheme such that $j_1$ is first coupled to $j_2$ to give a resultant $a_2$, and then $a_2$ is coupled to $j_3$ to yield $a_3$, and so on. Here $\vec{a}\equiv\{a_2,\cdots,a_{n-1}\}$ denotes the set of the angular momenta appeared in the intermediate coupling. The normalized gauge-invariant (or gauge-variant) intertwiner corresponds to the resulting angular momentum $J=0$ (or $J\neq0$). For the convenience of graphical calculus considered in this paper, we specify the normalized gauge-invariant intertwiner associated to a vertex $v$ as
\begin{align}\label{gauge-invariant-intertwiner}
&(i^{\vec{a}}_v)_{m_1m_2\cdots m_n}:=(-1)^{2j_n}{\left(i^{\,J=0;\,\vec{a}}_v\right)_{m_1m_2\cdots m_n}}^{M=0}\notag\\
=&(-1)^{2j_n}\prod_{i=2}^{n-1}\sqrt{d_{a_i}}\;
\makeSymbol{\includegraphics[width=5.4cm]{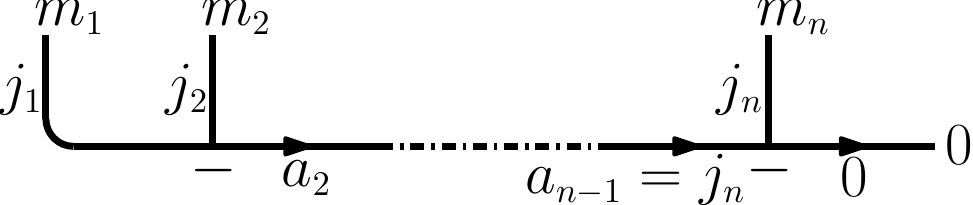}}\notag\\
=&\prod_{i=2}^{n-2}\sqrt{d_{a_i}}\;\makeSymbol{\includegraphics[width=5.7cm]{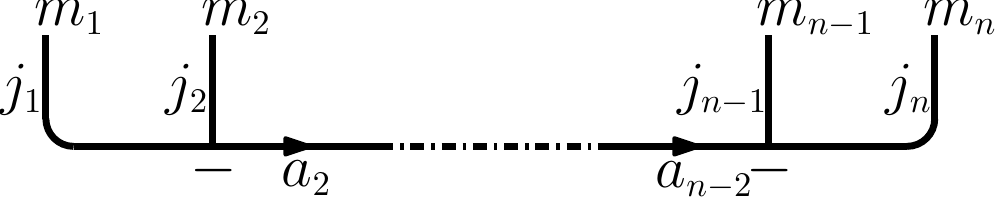}},
\end{align}
where in the second step the identities \eqref{arrow-3j} and \eqref{two-arrow-result} (see  below) were used. Hence the $3j$-symbol \eqref{3j-def-graph} is indeed the normalized gauge-invariant intertwiner associated to a trivalent vertex $v$ from which three edges start. The ``metric" tensor \eqref{metric-graph}, as the special $3j$-symbol, is the gauge-invariant intertwiner associated to a divalent vertex $v$ from which two edges start. It can be normalized by multiplying a factor $1/\sqrt{d_j}$.  The Kronecker delta \eqref{Kronecker-graph} in ${\cal H}_j$ is the gauge-invariant intertwiner associated to a divalent (trivial) vertex $v$ such that $v=b(e)=f(e')$ is the intersection of two edges $e$ and $e'$, which can be normalized by multiplying a factor $1/\sqrt{d_j}$.

Now we turn to the graphical transformation rules reflecting the properties of the $3j$-symbol. The $3j$-symbol has the following cyclic symmetries. An even permutation of the columns of the $3j$-symbol keeps its value unchanged, while an odd permutation leads to a multiplication by a factor $(-1)^{j_1+j_2+j_3}$, i.e.,
\begin{align}\label{3j-orientation-change-graph}
\makeSymbol{
\includegraphics[width=2cm]{figures/graphical-rules/wigner-3j-symbol-1}}&=(-1)^{j_1+j_2+j_3}\makeSymbol{
\includegraphics[width=2cm]{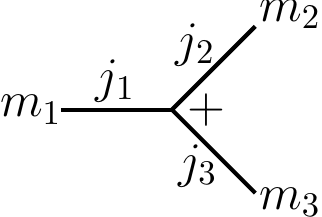}}\notag\\
&=(-1)^{j_1+j_2+j_3}\makeSymbol{
\includegraphics[width=2cm]{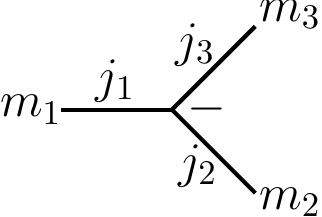}}\,.
\end{align}
The two orthogonality relations for $3j$-symbols are represented by the graphical rules
\begin{align}\label{3j-orthogonality-1}
\sum_jd_j\makeSymbol{
\includegraphics[width=1.2cm]{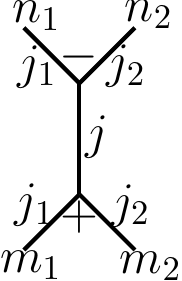}}&=\makeSymbol{
\includegraphics[width=1.2cm]{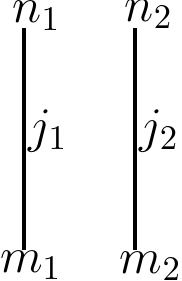}},\\
\makeSymbol{
\includegraphics[width=0.9cm]{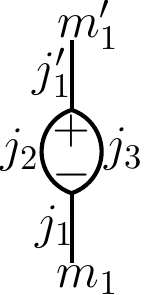}}&=\frac{\delta_{j_1,j'_1}}{d_{j_1}}\makeSymbol{
\includegraphics[width=0.5cm]{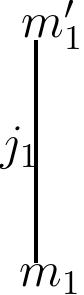}}.\label{3j-orthogonality-2}
\end{align}
From Eqs. \eqref{Kronecker-graph} and \eqref{3j-orthogonality-2}, one can easily obtain the following two graphical rules
\begin{align}
\makeSymbol{
\includegraphics[width=1.4cm]{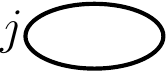}}&=d_j,\label{circle-rule}\\
\makeSymbol{
\includegraphics[width=2cm]{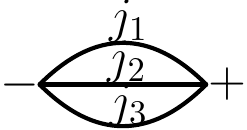}}&=1.\label{theta-rule}
\end{align}
The rules of reversing, removing and adding arrows in a graph read
\begin{align}
\makeSymbol{
\includegraphics[width=2cm]{figures/graphical-rules/wigner-3j-symbol-1}}&=\makeSymbol{
\includegraphics[width=2cm]{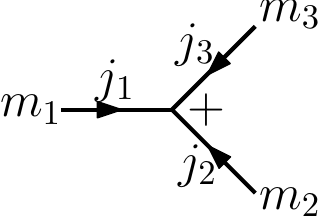}}=\makeSymbol{
\includegraphics[width=2cm]{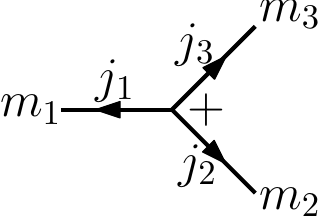}}\,,\label{three-arrow-adding}\\
\makeSymbol{
\includegraphics[width=2cm]{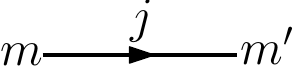}}&=(-1)^{2j}\makeSymbol{
\includegraphics[width=2cm]{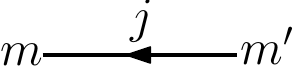}}\,,\label{arrow-flip}\\
\makeSymbol{
\includegraphics[width=2cm]{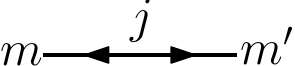}}&=
\makeSymbol{
\includegraphics[width=2cm]{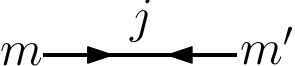}}=\makeSymbol{
\includegraphics[width=2cm]{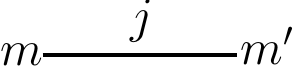}}\,,\label{two-arrow-cancel}\\
\makeSymbol{
\includegraphics[width=2cm]{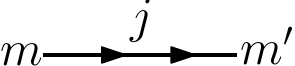}}&=\makeSymbol{
\includegraphics[width=2cm]{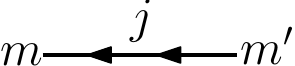}}=(-1)^{2j}\makeSymbol{
\includegraphics[width=2cm]{figures/graphical-rules/wigner-3j-symbol-9}}\,.\label{two-arrow-result}
\end{align}
These and the following rules are also useful to simplify graphs. A graph is regarded as a block diagram with $n$ external lines if, by rules \eqref{three-arrow-adding}--\eqref{two-arrow-result}, it can be transformed into the form such that every internal line has exactly one arrow and every external line has no arrow. Then the block diagram can be decomposed for different $n$ as follows.
\begin{enumerate}[(a)]
\item $n=1$
 \begin{align}\label{block-1}
 &\makeSymbol{\includegraphics[width=2.6cm]{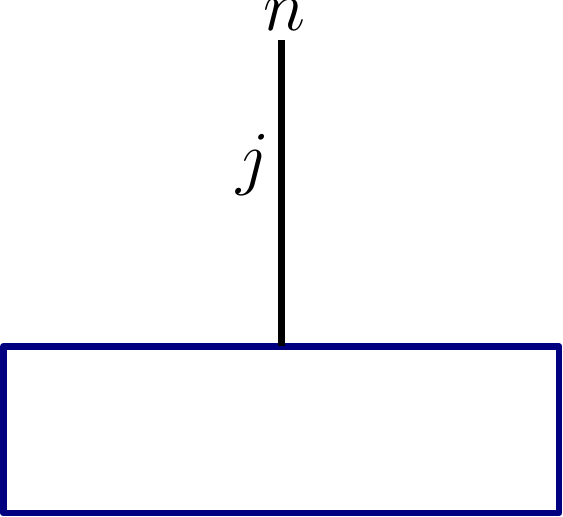}}=\delta_{j,0}\delta_{n,0}\;\makeSymbol{\includegraphics[width=2.6cm]{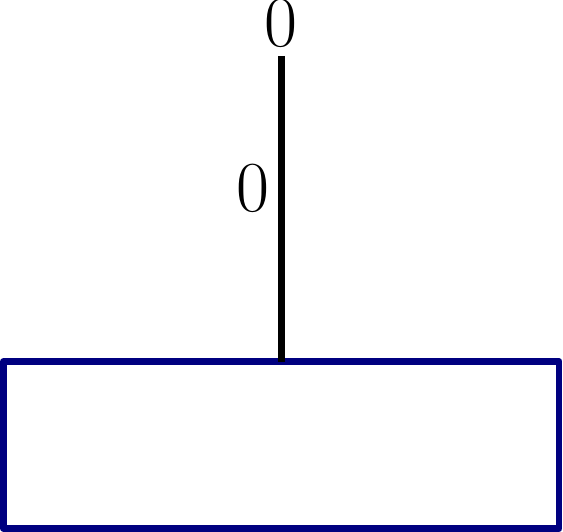}}.
\end{align}
\item $n=2$
 \begin{align}\label{block-2}
 &\makeSymbol{\includegraphics[width=2.6cm]{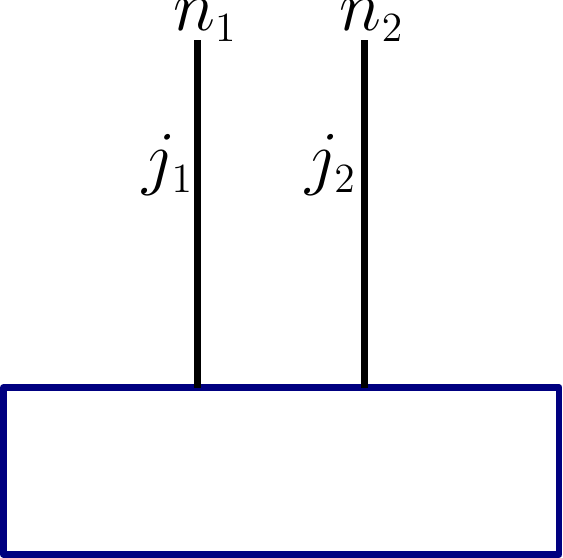}}=\frac{\delta_{j_1,j_2}}{d_{j_1}}\;\makeSymbol{\includegraphics[width=2.6cm]{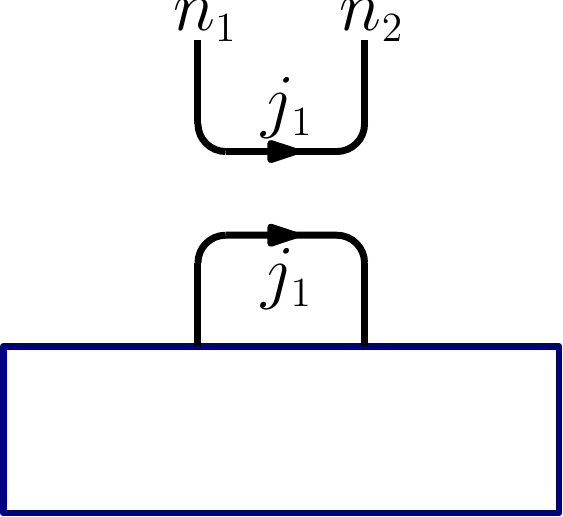}}.
\end{align}
 \item $n=3$
 \begin{align}\label{block-3}
 &\makeSymbol{\includegraphics[width=2.6cm]{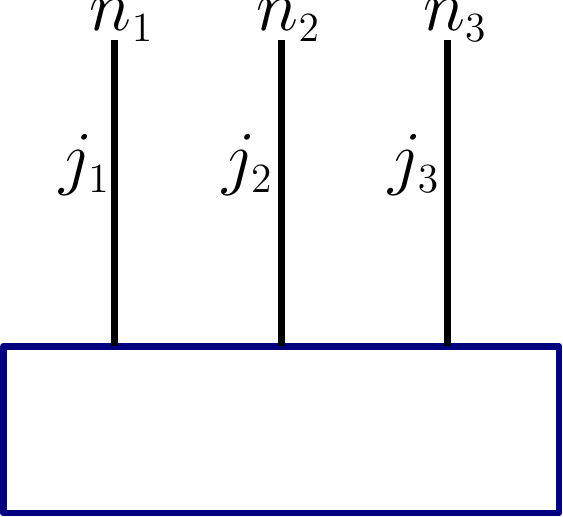}}=\;\makeSymbol{\includegraphics[width=2.6cm]{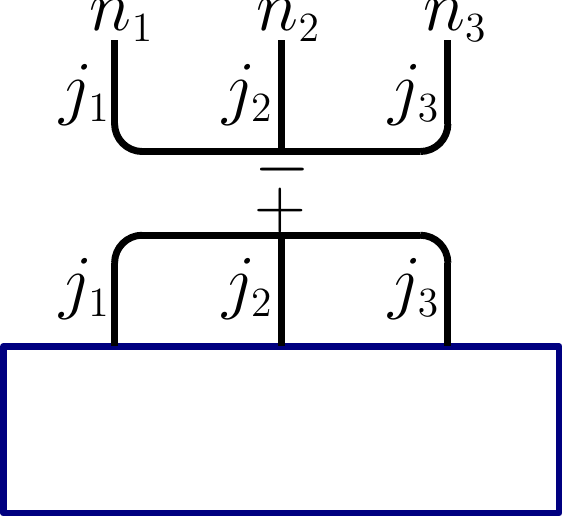}}.
\end{align}
\item $n>3$
\begin{align}\label{block-4}
 &\makeSymbol{\includegraphics[width=5.5cm]{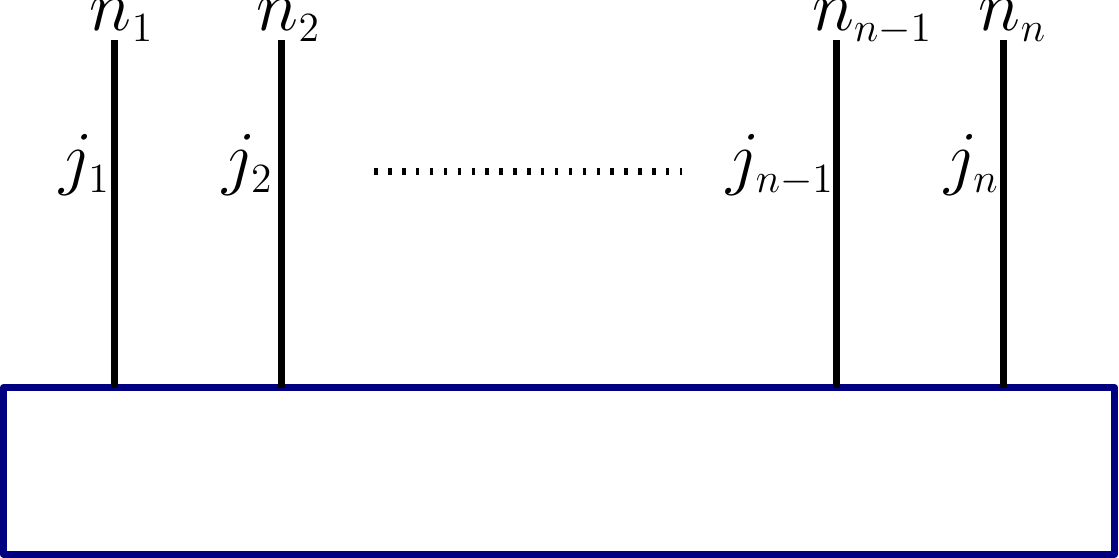}}\notag\\
 =&\sum_{\{a_2,\cdots,a_{n-2}\}}\prod_{i=2}^{n-2}d_{a_i}\;\makeSymbol{\includegraphics[width=5.5cm]{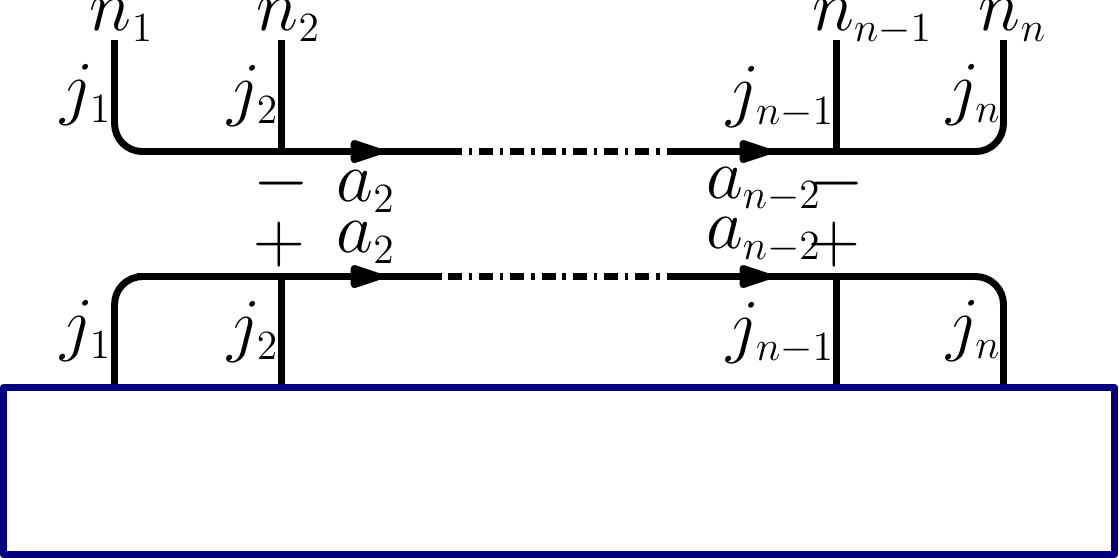}}.
\end{align}
\end{enumerate}
A direct application of Eq. \eqref{block-3} yields
\begin{align}\label{graph-simplify-related-6j}
\makeSymbol{
\includegraphics[width=3.2cm]{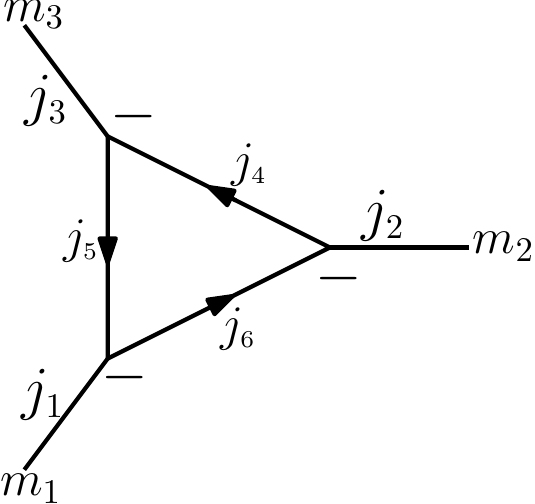}}=\makeSymbol{
\includegraphics[width=2cm]{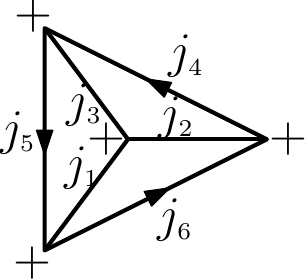}}\makeSymbol{
\includegraphics[width=2cm]{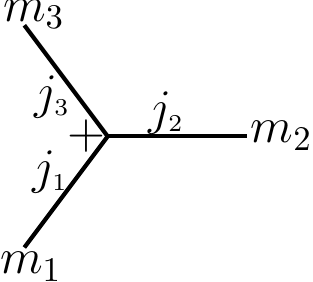}}\,,
\end{align}
where the first graph on the right-hand side represents a $6j$-symbol, i.e.,
\begin{align}\label{6j-def-graph}
\makeSymbol{
\includegraphics[width=2cm]{figures/graphical-rules/6j-symbol-def}}=\begin{Bmatrix}
j_1 & j_2 & j_3\\
j_4 & j_5 & j_6
\end{Bmatrix}\,,
\end{align}
which is obtained by contracting four $3j$-symbols. The $6j$-symbol is invariant by any permutation of columns and exchange of an upper and a lower arguments in each of column, e.g.,
\begin{align}\label{6j-symmetry-properties}
\begin{Bmatrix}
j_1 & j_2 & j_3\\
j_4 & j_5 & j_6
\end{Bmatrix}&=\begin{Bmatrix}
j_2 & j_1 & j_3\\
j_5 & j_4 & j_6
\end{Bmatrix}=\begin{Bmatrix}
j_3 & j_2 & j_1\\
j_6 & j_5 & j_4
\end{Bmatrix}=\cdots\notag\\
&=\begin{Bmatrix}
j_4 & j_5 & j_3\\
j_1 & j_2 & j_6
\end{Bmatrix}=\begin{Bmatrix}
j_4 & j_2 & j_6\\
j_1 & j_5 & j_3
\end{Bmatrix}=\cdots\,.
\end{align}
The matrix representation ${[\pi_j(g_e)]^m}_n$ of holonomy $g_{e}\equiv g_e(A)\in SU(2)$ of a $su(2)$-valued connection $A$ along an edge $e$ on $\Sigma$ is denoted by a colored line (not a black line) with an arrow on it as \cite{Yang:2015wka}\begin{align}\label{rep-group-graph}
{[\pi_j(g_e)]^m}_n=\;\makeSymbol{
\includegraphics[width=2.4cm]{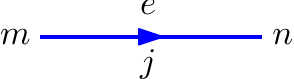}}\,.
\end{align}
All the information about ${[\pi_j(g_e)]^m}_n$ have been encoded in the graph in the right hand side of Eq. \eqref{rep-group-graph}. The corresponding irreducible representation $\pi_j$ of $g_e$ is denoted by $e$ and $j$ labeling the line. The orientation of $e$ with respect to the vertices is reflected by the orientation of the arrow on the line. The row (former or up) index and the column (latter or down) index are denoted by the two indices $m$ and $n$ labeling the starting and the ending points of the line, respectively, and the orientation of the arrow is from its row index $m$ to its column index $n$. The graphical transformations for the holonomy consist of the following two rules. First, a transformation from the irreducible representation of $g^{-1}_e$ to that of $g_e$ is given by
\begin{align}\label{rep-inverse-graph}
{[\pi_j(g_e^{-1})]^n}_{\,m}&={[\pi_j(g_{e^{-1}})]^n}_{\,m}=C^{(j)}_{mm'}\,{[\pi_j(g_e)]^{m'}}_{n'}C^{n'n}_{(j)}\notag\\
&=\makeSymbol{
\includegraphics[width=2.4cm]{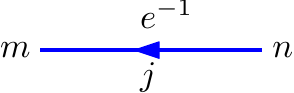}}=\makeSymbol{
\includegraphics[width=3.6cm]{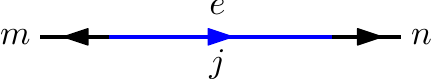}}\,.
\end{align}
Second, coupling two representations of the same holonomy $g_e$, corresponding to the Clebsch-Gordan series, is expressed as
\begin{align}\label{holonomy-reps-couple}
&{[\pi_{j_1}(g_e)]^{m_1}}_{\,n_1}{[\pi_{j_2}(g_e)]^{m_2}}_{\,n_2}=\makeSymbol{
\includegraphics[width=2.6cm]{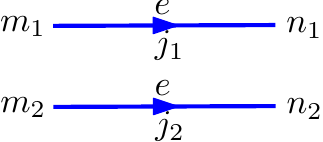}}\notag\\
=&\sum_{j_3}d_{j_3}\makeSymbol{
\includegraphics[width=4.4cm]{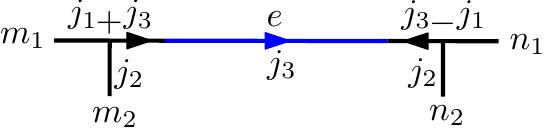}}\notag\\
=&\sum_{j_3}d_{j_3}\makeSymbol{
\includegraphics[width=4.4cm]{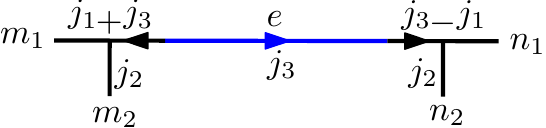}},
\end{align}
which can be easily generalized to
\begin{align}\label{recoupling-rule}
&{[\pi_{j_1}(g_e)]^{m_1}}_{\,n_1}{[\pi_{j_2}(g_e)]^{m_2}}_{\,n_2}\cdots{[\pi_{j_n}(g_e)]^{m_n}}_{\,n_n}\notag\\
=&\makeSymbol{
\includegraphics[width=5.5cm]{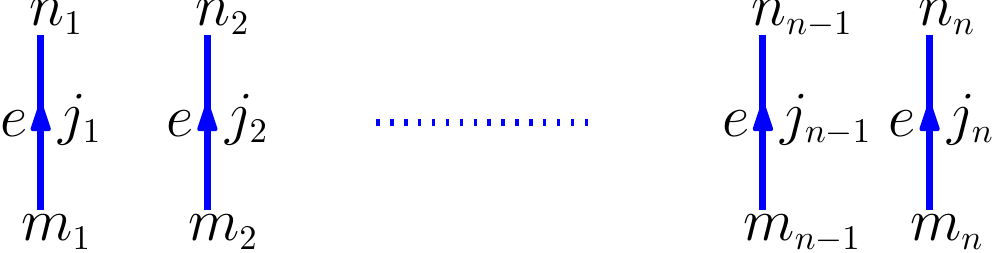}}\notag\\
=&\sum_{a_2}d_{a_2}\makeSymbol{
\includegraphics[width=4.5cm]{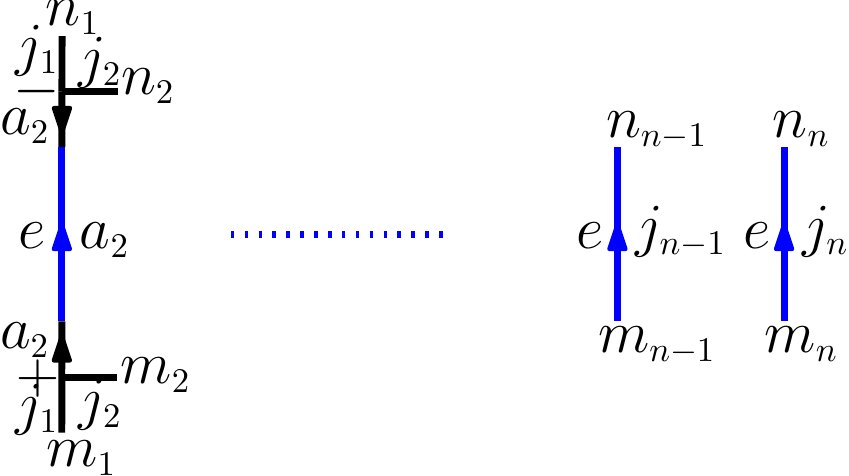}}\notag\\
=&\sum_{a_2}d_{a_2}\makeSymbol{
\includegraphics[width=5.5cm]{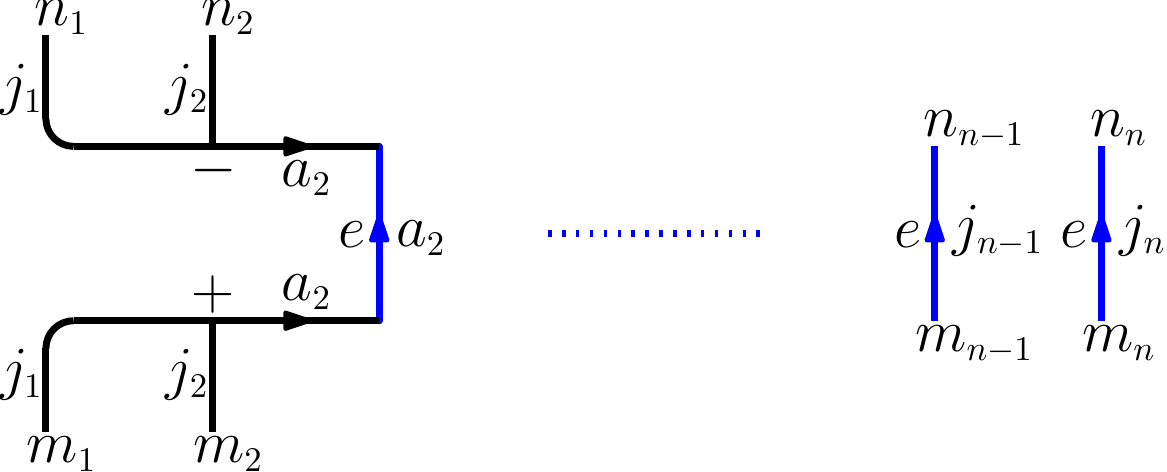}}\notag\\
=&\sum_{\{a_2,\cdots,a_{n-1},J\}}\prod_{i=2}^{n-1}d_{a_i}d_J\makeSymbol{
\includegraphics[width=5.5cm]{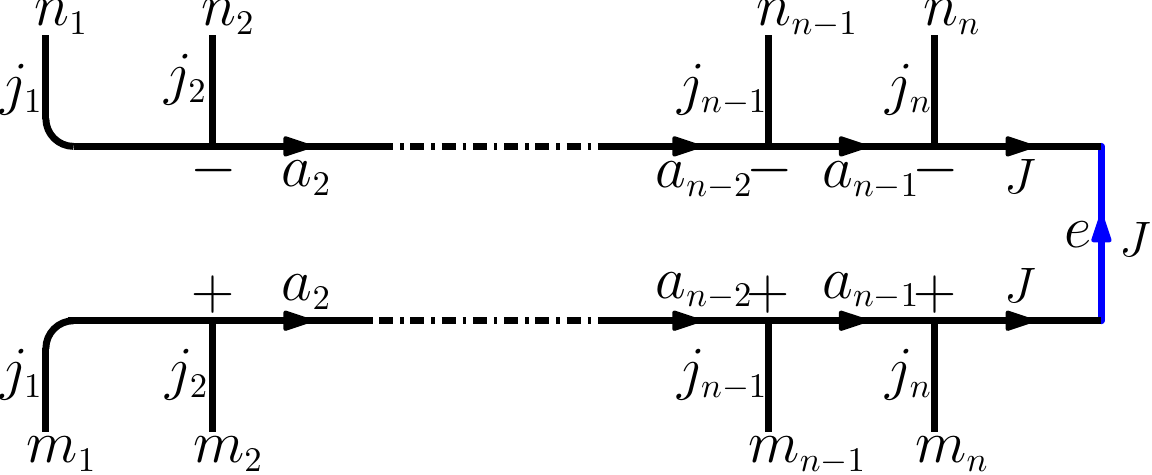}}.
\end{align}

Now we extend the above graphical calculus to compute the integral over the product of irreducible representations of $g_e$. By Eq. \eqref{recoupling-rule}, the integral can be evaluated graphically by
\begin{align}\label{integral-graph-1}
&\int{\rm d}g_e{[\pi_{j_1}(g_e)]^{m_1}}_{\,n_1}{[\pi_{j_2}(g_e)]^{m_2}}_{\,n_2}\cdots{[\pi_{j_n}(g_e)]^{m_n}}_{\,n_n}\notag\\
=&\int{\rm d}g_e\makeSymbol{
\includegraphics[width=5.5cm]{figures/graphical-rules/sfm-gra-1}}\notag\\
=&\sum_{\{a_2,\cdots,a_{n-1},J\}}\prod_{i=2}^{n-1}d_{a_i}d_J\notag\\
&\quad\times\int{\rm d}g_e\makeSymbol{
\includegraphics[width=5.7cm]{figures/graphical-rules/sfm-gra-4}}\notag\\
=&\sum_{\{a_2,\cdots,a_{n-2}\}}\prod_{i=2}^{n-2}d_{a_i}\makeSymbol{
\includegraphics[width=5cm]{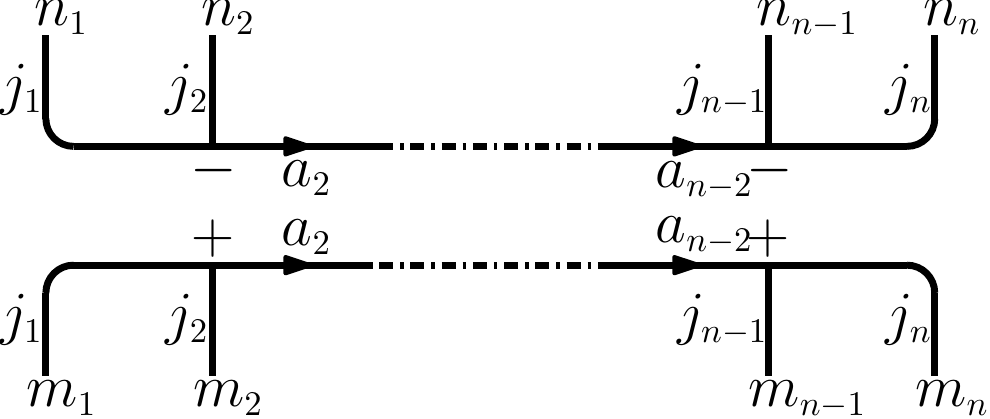}}\notag\\
=&\sum_{\vec a}(i^{\vec{a}}_{v_1})^{m_1\cdots m_n}(i^{\vec{a}}_{v_2})_{n_1\cdots n_n},
\end{align}
where we used $\int {\rm d}g_e{[\pi_{J}(g_e)]^{M}}_{\,N}=\delta_{J,0}\delta^{M,0}\delta_{N,0}$, the identities \eqref{arrow-3j}, \eqref{two-arrow-result} and $(-1)^{4j_n}=1$ in the third step. Thus the integration over all group elements associated to $e$ gives two normalized gauge-invariant intertwiners $i_{v_1}$ and $i_{v_2}$ to the two vertices (endpoints) $v_1$ and $v_2$ of $e$. Similarly, one has the following useful formulas.
\begin{enumerate}[(a)]
\item $n=2$
\begin{align}
 &\int{\rm d}g_e\makeSymbol{
\includegraphics[width=1.5cm]{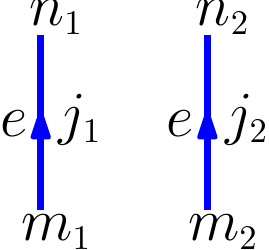}}=\frac{\delta_{j_1,j_2}}{d_{j_1}}\makeSymbol{
\includegraphics[width=1.3cm]{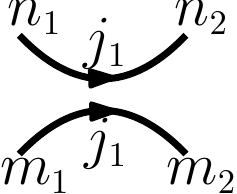}}=\frac{\delta_{j_1,j_2}}{d_{j_1}}\makeSymbol{
\includegraphics[width=1.3cm]{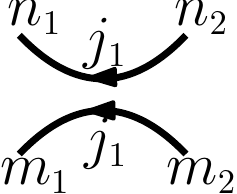}},\label{2-edges-integral-1}\\
 &\int{\rm d}g_e\makeSymbol{
\includegraphics[width=1.5cm]{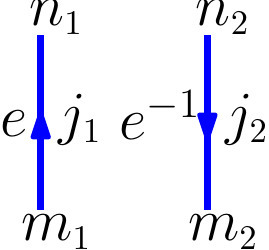}}=\frac{\delta_{j_1,j_2}}{d_{j_1}}\makeSymbol{
\includegraphics[width=1.3cm]{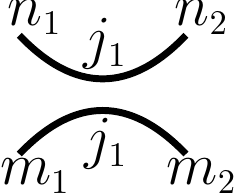}}\label{2-edges-integral-2}.
\end{align}
\item $n=3$
\begin{align}
 \int{\rm d}g_e\makeSymbol{
\includegraphics[width=2.4cm]{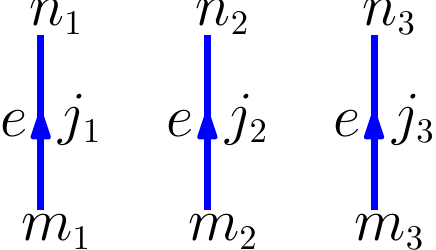}}=\makeSymbol{
\includegraphics[width=2.3cm]{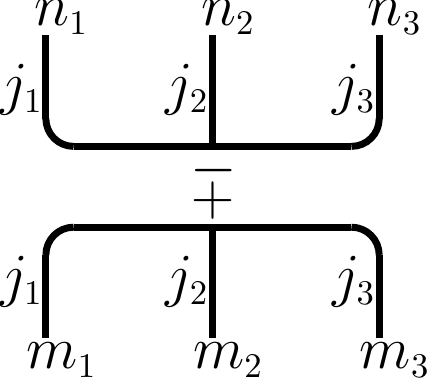}}\label{2-edges-integral}.
\end{align}
\end{enumerate}
In SFMs, one usually needs also to evaluate the integration over the product of irreducible representations of $g_e$ and its inverse $g^{-1}_e=g_{e^{-1}}$. This can be accomplished by combining Eq. \eqref{integral-graph-1} with Eq. \eqref{rep-inverse-graph}. For example, the integration over the product of $n-1$ irreducible representations of $g_e$ and one representation of $g^{-1}_e$ is given graphically by
\begin{align}\label{graph-int-explicit}
&\int{\rm d}g_e\makeSymbol{
\includegraphics[width=5.5cm]{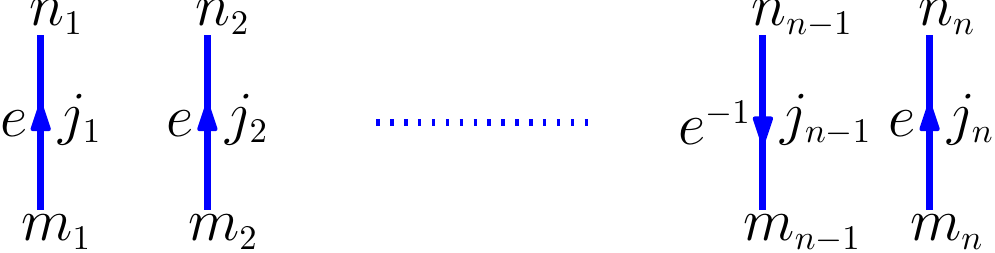}}\notag\\
=&\int{\rm d}g_e\makeSymbol{
\includegraphics[width=5.5cm]{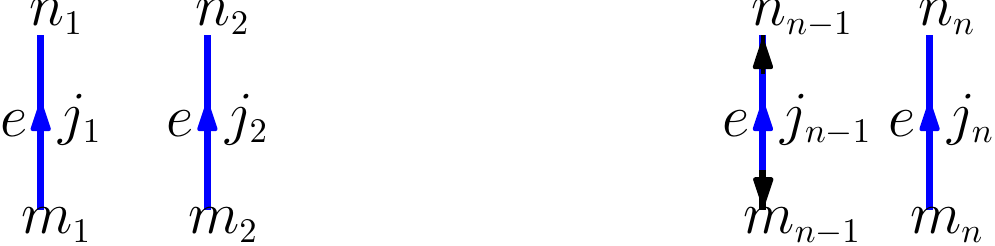}}\notag\\
=&\sum_{\{a_2,\cdots,a_{n-2}\}}\prod_{i=2}^{n-2}d_{a_i}\makeSymbol{
\includegraphics[width=5cm]{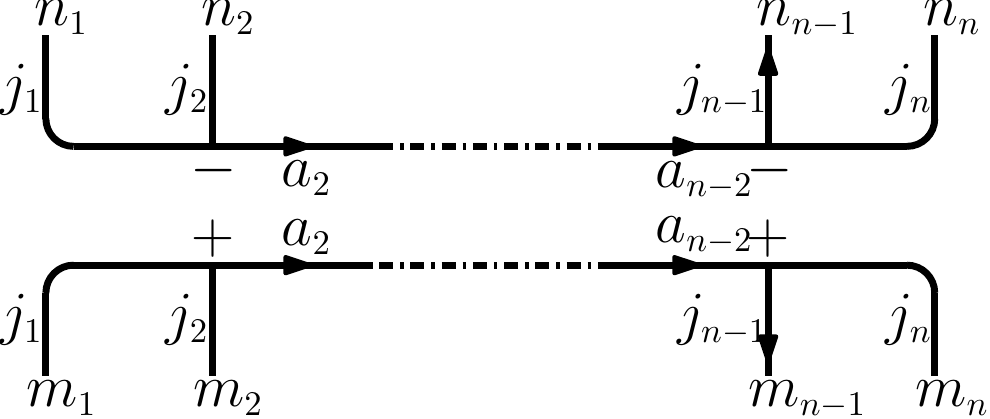}}.
\end{align}
Thus the integration leads to a summation of products of (normalized) two intertwiners $i_{v_1}$ and $i_{v_2}$ associated to two endpoints $v_1$ and $v_2$ of $e$. Each of the intertwiners has an arrow on the external line associated to $g^{-1}_e$ while the other external lines have no arrows. By the transformation rules \eqref{three-arrow-adding}-\eqref{two-arrow-result}, the above result can be transformed to an equivalent form such that each external lines associated to $g_e$ has an arrow while the external line associated to $g^{-1}_e$ has no arrow. In the case that the orientations of edges are irrelevant to the question considered, the corresponding integration can be roughly expressed as
\begin{align}\label{graph-int-rough}
&\int{\rm d}g_e\makeSymbol{
\includegraphics[width=5.5cm]{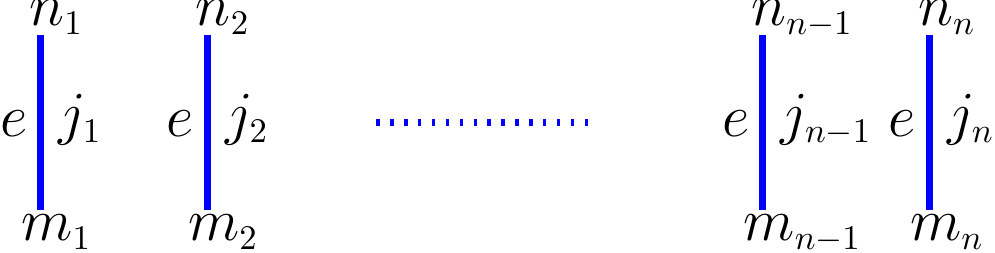}}\notag\\
=&\makeSymbol{
\includegraphics[width=5cm]{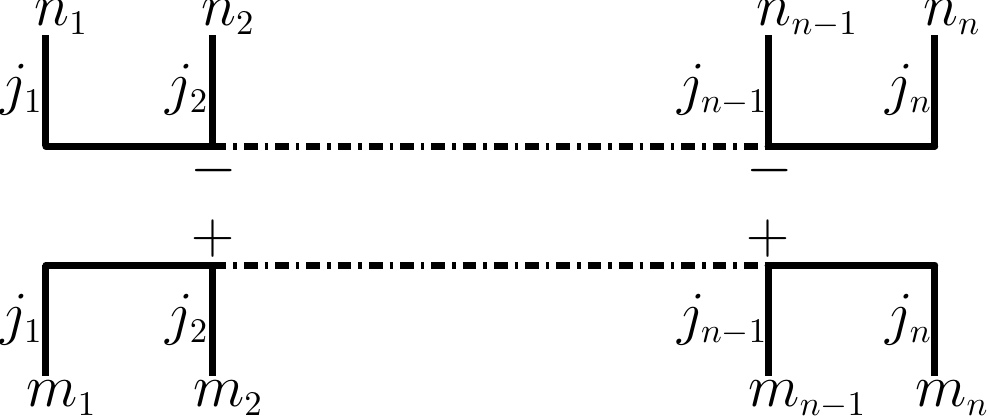}}.
\end{align}
We will adopt the rough formula \eqref{graph-int-rough} to derive the partition function in the following two subsections. Certain explicit formula with orientations of edges similar to \eqref{graph-int-explicit} will be considered in Sec. \ref{sec-IV}.

\subsection{The partition function in nonboundary cases}
Let us consider the underlying dual 2-cell $\Delta^*$ without boundary, and derive the partition function on $\Delta^*$ from Eq. \eqref{Z-BF-alg} by the graphical calculus presented in the previous subsection. In nonboundary cases, the partition function \eqref{Z-BF-alg} reduces to
\begin{align}\label{BF-partition-function-Af}
 {\cal Z}^{\rm BF}(\Delta^*)&=\int{\rm d}g^+_{fv}\prod_{f\in\Delta^*}\sum_{j^+_f}d_{j^+_f}{\rm Tr}_{j^+_f}\left(\prod_{v\in\partial f}g^+_{fv}\right)\notag\\
 &\hspace{1.4cm}\times\prod_{fv}\int {\rm d}g^+_{ve}\sum_{j^+_{fv^{-1}}}d_{j^+_{fv^{-1}}}{\rm Tr}_{j^+_{fv^{-1}}}(g^+_{e'v}g^+_{ve}g^+_{fv^{-1}})\notag\\
 &\quad\times\int{\rm d}g^-_{fv}\prod_{f\in\Delta^*}\sum_{j^-_f}d_{j^-_f}{\rm Tr}_{j^-_f}\left(\prod_{v\in\partial f}g^-_{fv}\right)\notag\\
 &\hspace{1.4cm}\times\prod_{fv}\int {\rm d}g^-_{ve}\sum_{j^-_{fv^{-1}}}d_{j^-_{fv^{-1}}}{\rm Tr}_{j^-_{fv^{-1}}}(g^-_{e'v}g^-_{ve}g^-_{fv^{-1}})\notag\\
 &\equiv\int{\rm d}g^+_{fv}{\rm d}g^-_{fv}\prod_{f\in\Delta^*}A^{\rm BF}_f(\{g^+_{fv},g^-_{fv}\}).
\end{align} 
To derive a partition function $Z^{\rm SFM}(\Delta^*)$ from \eqref{BF-partition-function-Af}, one can follow the three steps introduced below Eq. \eqref{eq:delta-delta}.

First, we need to perform the ${\rm d}g^+_{ve}{\rm d}g^-_{ve}$ integration in the first equality in Eq. \eqref{BF-partition-function-Af} to obtain the expression of $A^{\rm BF}_f$. Notice that a segment $ve$ bounded by $n$ faces contributes the partition function \eqref{BF-partition-function-Af} $n$ pairs ($g^+_{ve},g^-_{ve}$) of matrix elements to be integrated out. To perform the integration, we first transform the algebraic formula \eqref{BF-partition-function-Af} into its graphical formula. Thanks to the graphical calculus presented in Eq. \eqref{graph-int-rough}, the ${\rm d}g^+_{ve}{\rm d}g^-_{ve}$ integration can be straightforwardly evaluated to yield
\begin{align}
A^{\rm BF}_f(\{g^+_{fv},g^-_{fv}\})&=\makeSymbol{
\includegraphics[width=4.7cm]{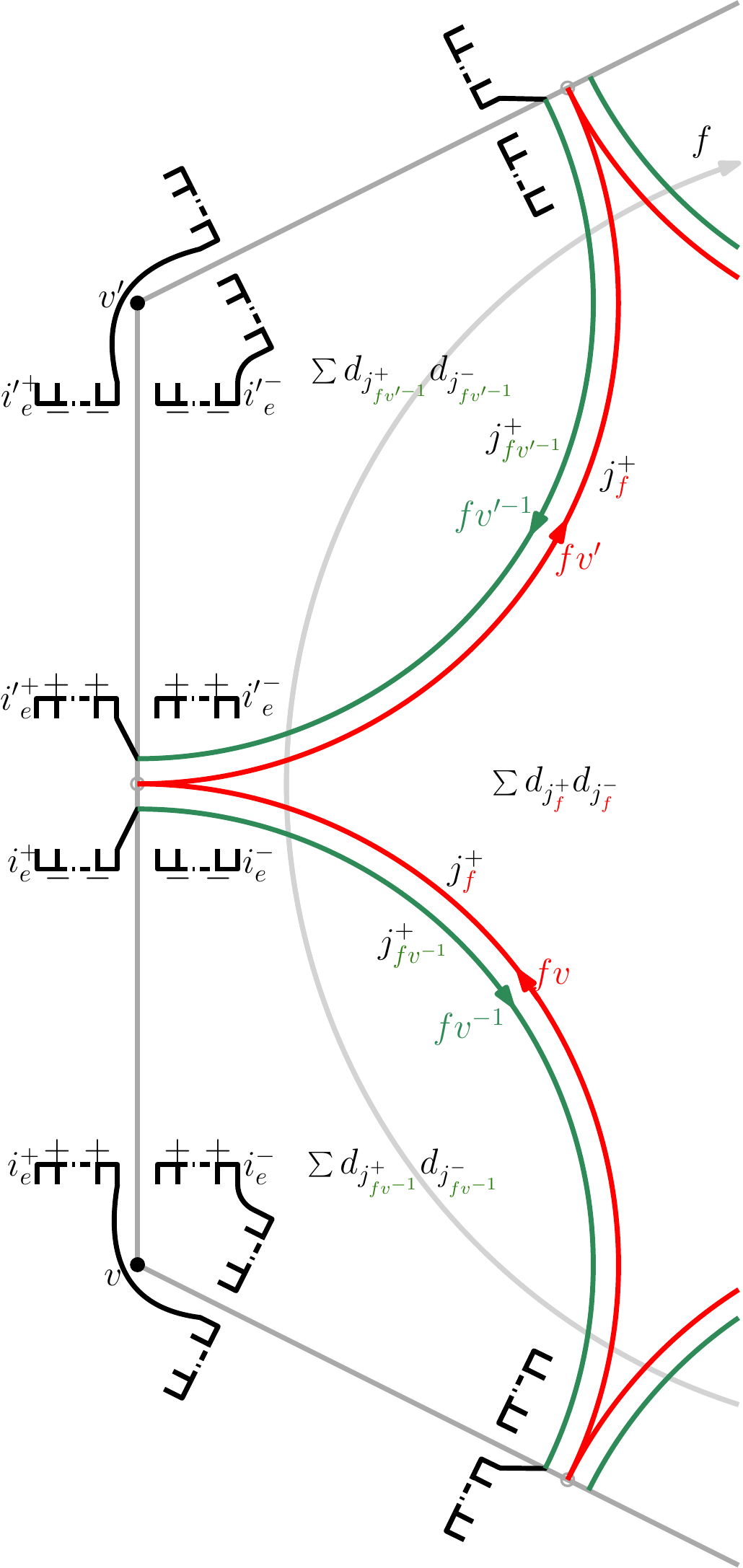}},
\end{align}
where the oriented red and dark green curved lines denote respectively the matrix elements of $g^+$ attached to edges $fv$ of $\gamma_v$ and their inverses $fv^{-1}$, while those of $g^{-}$ are omitted for simplicity. It should be noted that the pairs $({i'}^{+}_e,i^+_e)$ of intertwiners, as well as $({i'}^{-}_e,i^-_e)$, associated to the endpoint pair $(v',v)$ have certain arrows (the ``metrics") with uniform orientations on their external lines, while the intertwiners associated to the midpoints of $e$ have no arrows on their external lines. The origin of these arrows is the following. For a given internal edge $e$ bounded by faces $f\in\partial e$, each $f$ induces an orientation on $e$ and contributes a holonomy with representation $j_f$. If the induced orientations on $e$ from different faces are not the same, the induced holonomies with spins $j_f$ will involve $g_e$ as well as $g_{e^{-1}}$ associated to $e$. However, the representations of $g_{e^{-1}}$ can be uniformly transformed to those of $g_e$ by adding two arrows to the graphical representation of the corresponding hononomy associated to the endpoints $v$ and $v'$ of $e$ by the graphical rule \eqref{rep-inverse-graph}.

Second, we need to impose the simplicity constraint. In the Euclidean EPRL model as well as its generalized model, the simplicity constraint was first expressed as the corresponding linear formulation and then imposed at quantum level by the master-constraint criterion \cite{Engle:2007mu,Engle:2007wy} or the Gupta-Bleuler criterion \cite{Ding:2009jq}. The result restricts the relation between $j^\pm$ and their coupling $j$ associated to $f$, depending on the values of the Immirzi parameter $\beta$, as \cite{Engle:2007wy,Kaminski:2009fm,Ding:2010fw,Perez:2012wv}
\begin{align}
 \begin{cases}
   j^\pm=(1\pm\beta)j/2, \quad \text{for }\beta<1\\
   j^\pm=(\beta\pm1)j/2, \quad \text{for }\beta>1
 \end{cases},
\end{align}
and thus
\begin{align} 
 \begin{cases}
   j=j^++j^-, \quad \text{for }\beta<1\\
   j=j^+-j^-, \quad \text{for }\beta>1
 \end{cases}.
\end{align}
For the turning point $\beta=1$, one has
\begin{align}\label{eq:simplicity-beta-1}
 j^+=j,\quad j^-=0.
\end{align}
Hence the quantum simplicity constraint can be imposed as a projection by the so-called ${\cal Y}$ map
\begin{align}
{\cal Y}:{\cal H}_{j^+}\otimes {\cal H}_{j^-}=\oplus_{j'=|j^+-j^-|}^{j^++j^-}{\cal H}_{j'}\rightarrow{\cal H}_j.
\end{align}
It naturally induces a $Y$ map with actions on intertwiners graphically as
\begin{align}\label{Y-map-intertwiner}
Y\makeSymbol{
\includegraphics[width=1.4cm]{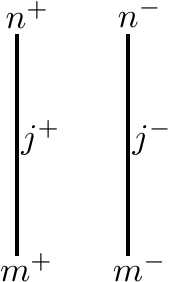}}=
 Y\sum_{j'=|j^+-j^-|}^{j^++j^-}d_{j'}\makeSymbol{
\includegraphics[width=1.4cm]{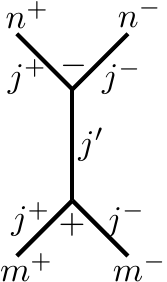}}:=d_j\makeSymbol{
\includegraphics[width=1.4cm]{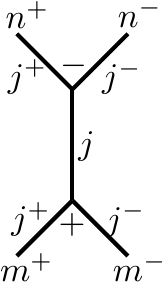}}.
\end{align}
By imposing the $Y$ map \eqref{Y-map-intertwiner} on the intertwiners associated to the vertices of $\gamma_v$, the partition function ${\cal Z}^{\rm BF}(\Delta^*)$ in Eq. \eqref{BF-partition-function-Af} can be promoted to the (Euclidean) generalized EPRL partition function
\begin{align}\label{Z-EPRL-A}
 {\cal Z}^{\rm EPRL}(\Delta^*):=\int{\rm d}g^+_{fv}{\rm d}g^-_{fv}\prod_{f\in\Delta^*}A^{\rm EPRL}_f(\{g^+_{fv},g^-_{fv}\})
\end{align}
with
\begin{align}
 A^{\rm EPRL}_f(\{g^+_{fv},g^-_{fv}\})&:=Y\cdot A^{\rm BF}_f(\{g^+_{fv},g^-_{fv}\})\notag\\
 &=\makeSymbol{
\includegraphics[width=4.7cm]{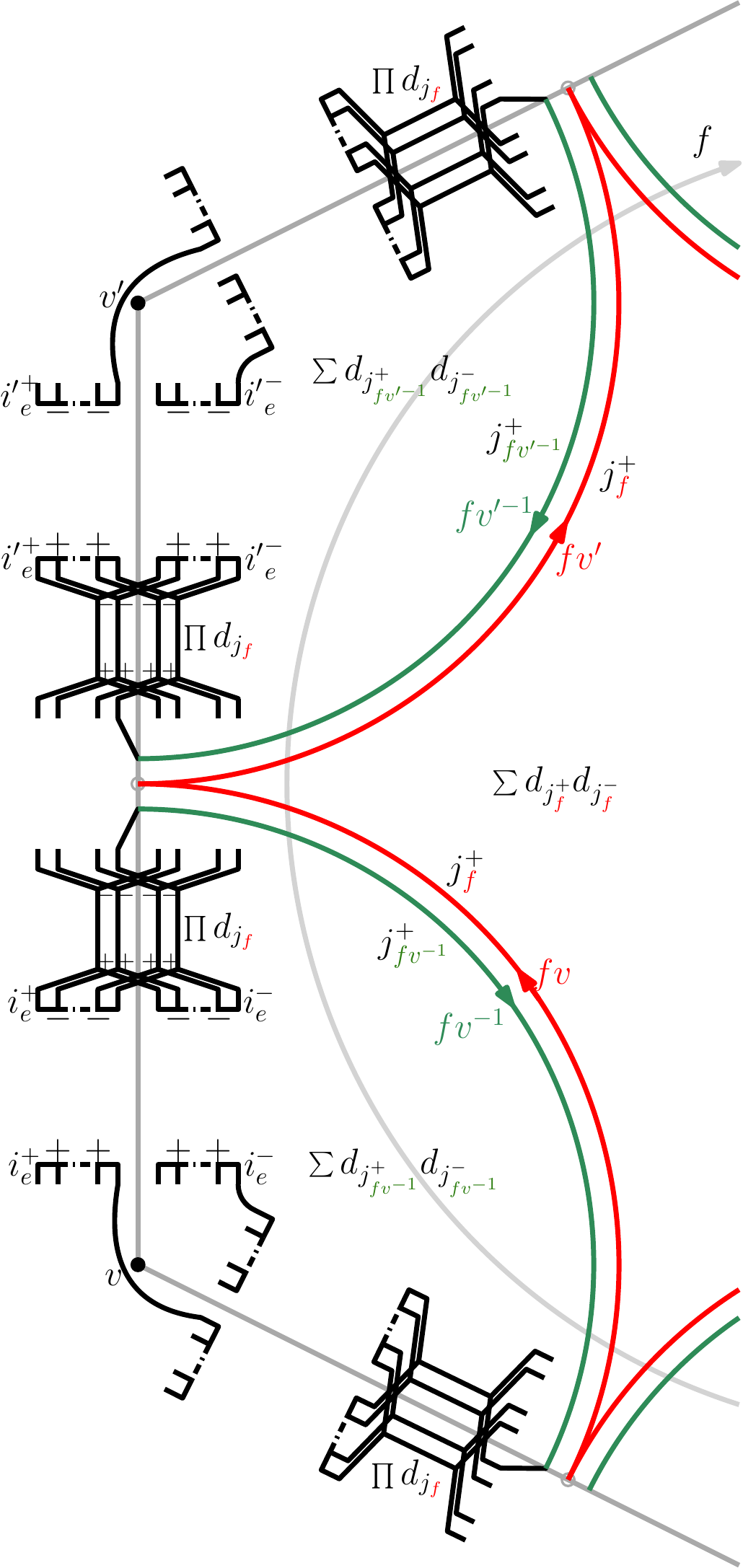}}\notag\\
&=\makeSymbol{
\includegraphics[width=4.7cm]{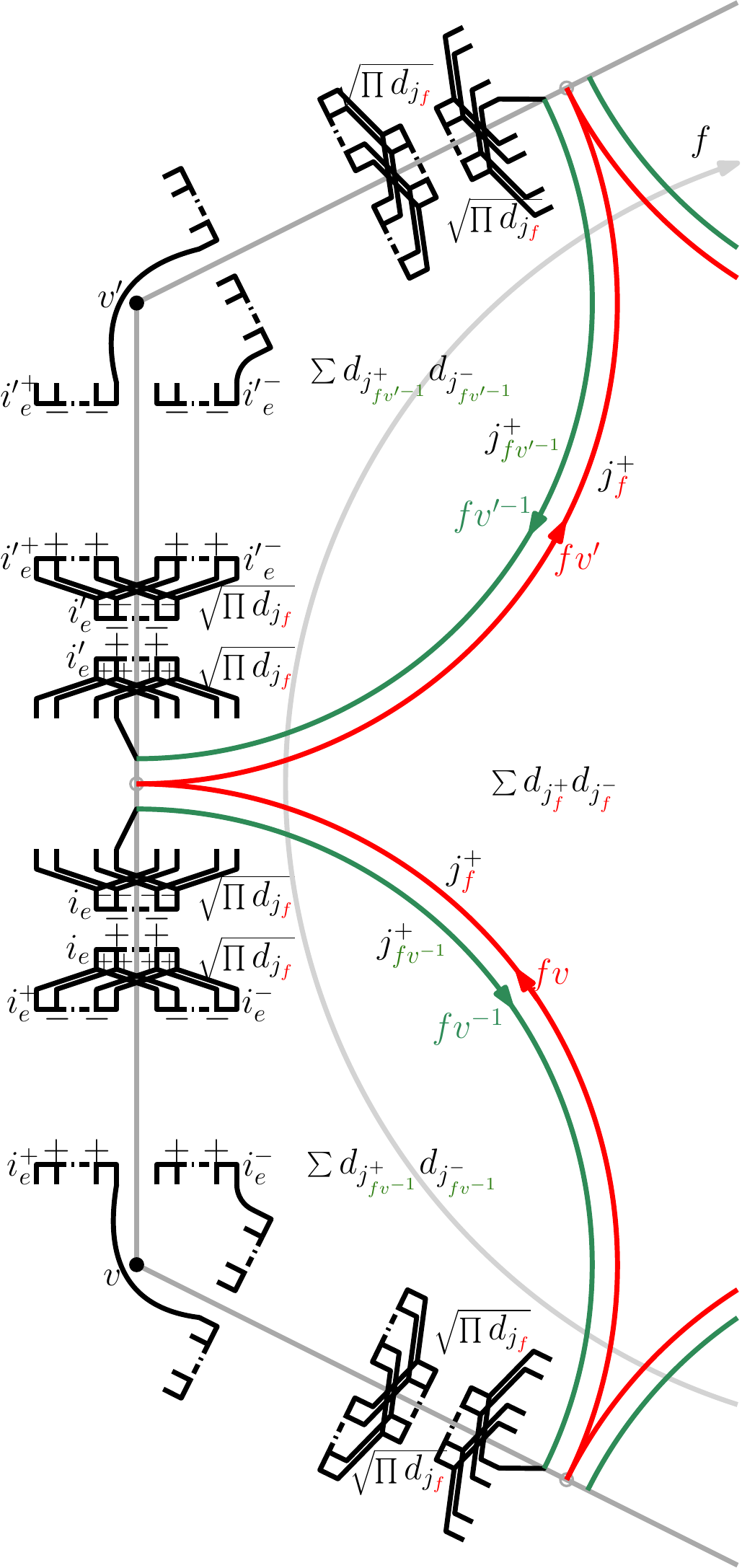}},
\end{align}
where Eqs. \eqref{Y-map-intertwiner} and \eqref{block-4} were used in the second and third steps respectively.

Third, we need to perform the integration over the group elements $g^\pm_{fv}$ associated to each vertex $v$. In the current case, Eq. \eqref{2-edges-integral-2} becomes
\begin{align}\label{graph-two-edges-integral}
\int {\rm d}g^+_{fv}\makeSymbol{
\includegraphics[width=2cm]{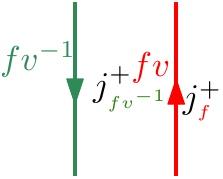}}&=\frac{\delta_{j^+_{fv^{-1}},j^+_f}}{d_{j^+_{fv^{-1}}}}\makeSymbol{
\includegraphics[width=1cm]{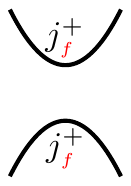}}.
\end{align}
By integration, Eq. \eqref{Z-EPRL-A} reduces to
\begin{align}\label{resulting-Z-nonboundary}
 {\cal Z}^{\rm EPRL}(\Delta^*)&=\prod_{f\in\Delta^*}\makeSymbol{
\includegraphics[width=4.7cm]{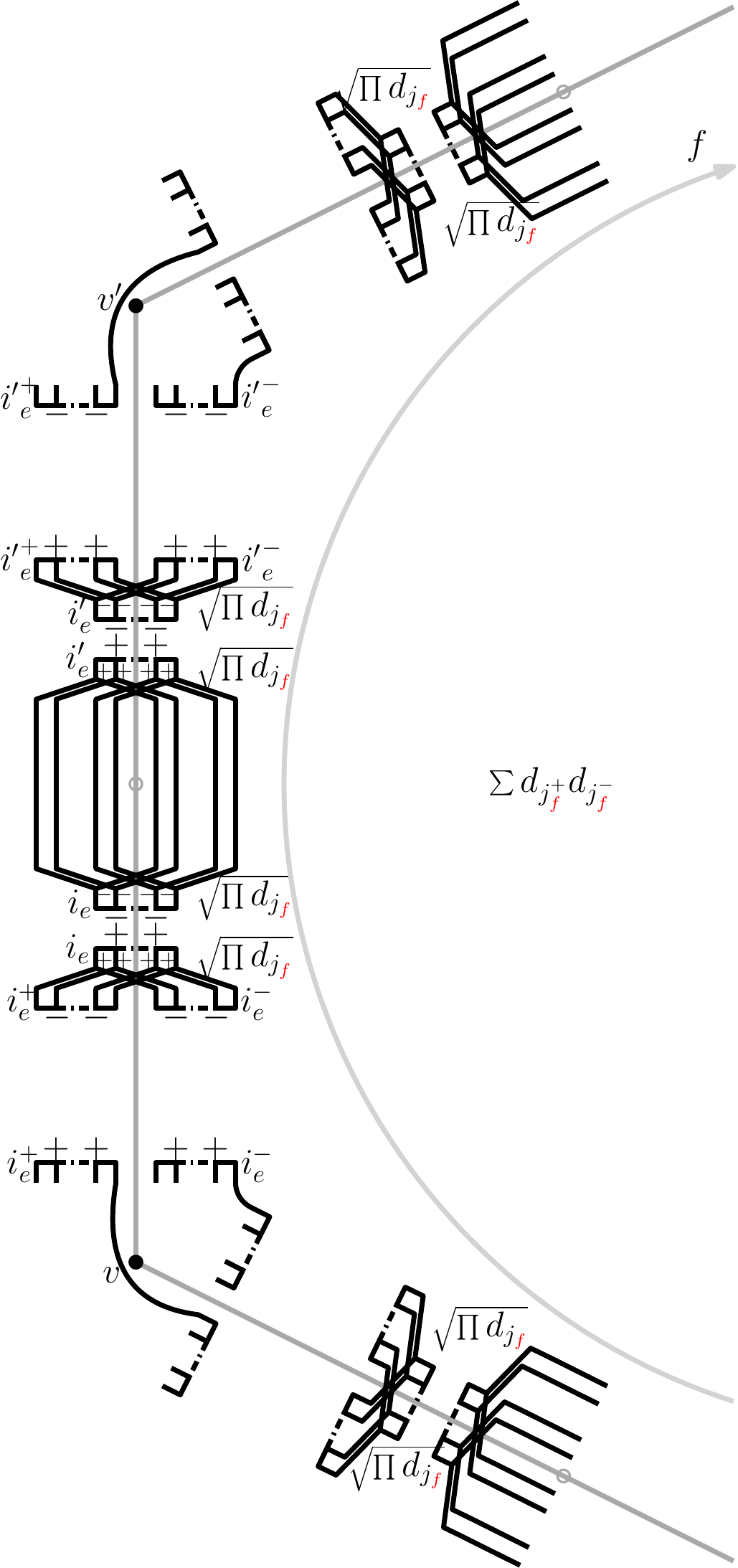}}\notag\\
&=\prod_{f\in\Delta^*}\makeSymbol{
\includegraphics[width=4.7cm]{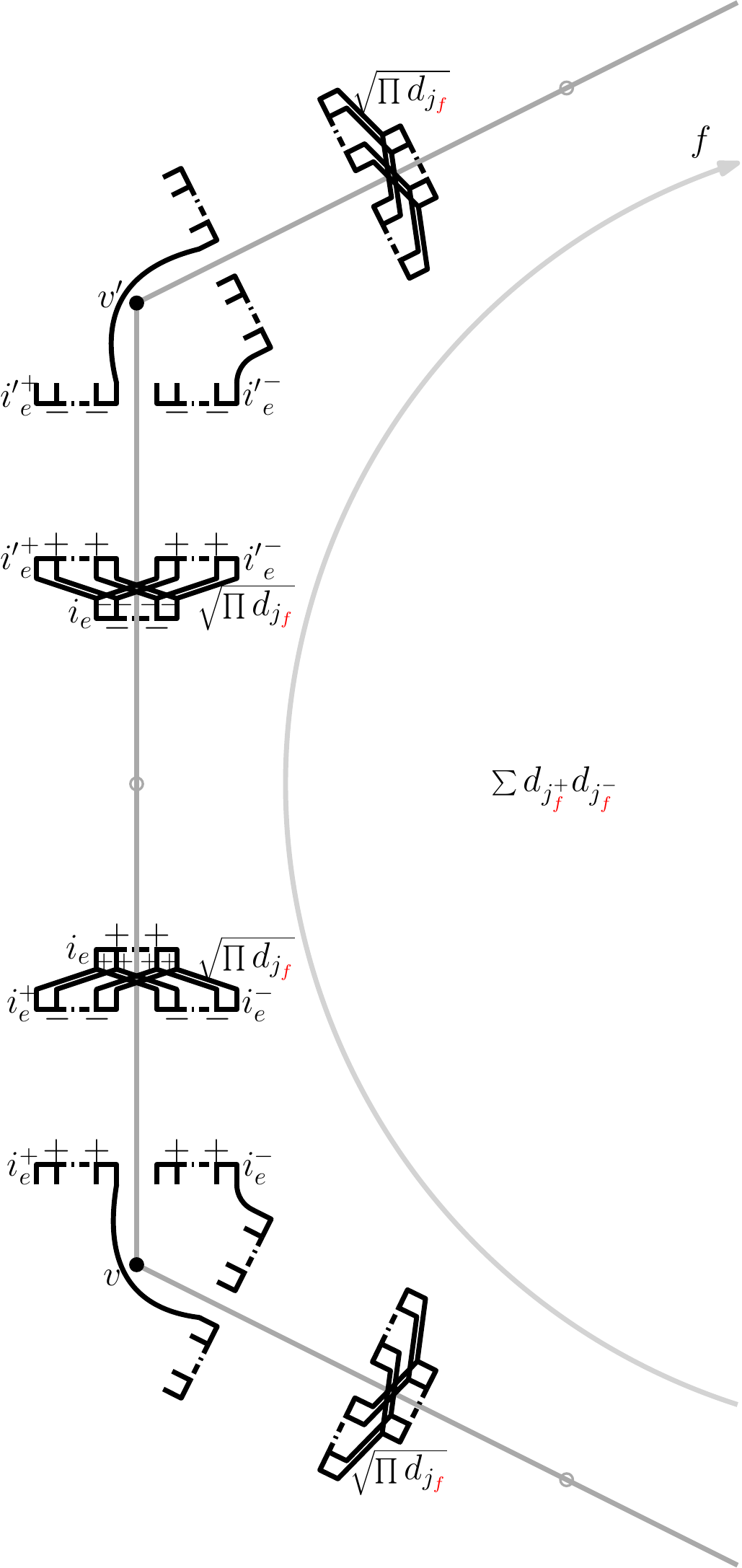}},
\end{align}
where in the second step we used
\begin{align}\label{circle-identity}
\makeSymbol{
\includegraphics[width=2.8cm]{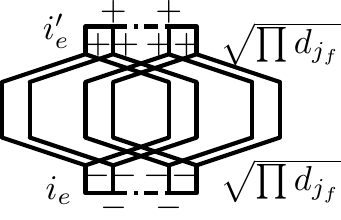}}&=\makeSymbol{
\includegraphics[width=1.4cm]{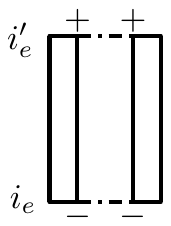}}=\delta_{i_e,i'_e},
\end{align}
in the light of Eq. \eqref{3j-orthogonality-2}. The resulting partition function ${\cal Z}^{\rm EPRL}(\Delta^*)$ in Eq. \eqref{resulting-Z-nonboundary} assigns to each internal vertex $v$ a contraction of intertwiners $i^+_e\otimes i^-_e$ associated to edges $e\in\partial v$, as a vertex amplitude $A_v$, to each internal edge $e$ a fusion coefficient
\begin{align}
f^{i_e}_{i^+_ei^-_e}\equiv \makeSymbol{
\includegraphics[width=3cm]{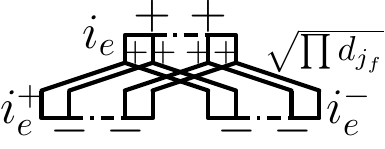}},
\end{align}
as an edge amplitude $A_e$, and to each face $f$ a factor $d_{j^+_f}d_{j^-_f}$ as a face amplitude $A_f$. The graphical formula of ${\cal Z}^{\rm EPRL}(\Delta^*)$ presented in Eq. \eqref{resulting-Z-nonboundary} can be uniquely transformed into its algebraic formula
\begin{align}\label{resulting-Z-nonboundary-algebraic}
 {\cal Z}^{\rm EPRL}(\Delta^*)&=\sum_{j^+_f,j^-_f}\prod_{f\in\Delta^*}d_{j^+_{f}}d_{j^-_{f}}\notag\\
 &\qquad\times\prod_{v\in\Delta^*}\sum_{i^+_e,i^-_e,i_e}{\rm Tr}_v\left[\bigotimes_{e\in\partial v}(i^+_e\otimes i^-_e)\right]\prod_{e\in\partial v} f^{i_e}_{i^+_ei^-_e},
\end{align}
which coincides with the one appeared in Refs. \cite{Ding:2010fw,Alesci:2011ia}.

\subsection{The partition function in cases with boundaries}
Now let us consider the underlying dual 2-cell $\Delta^*$ with a boundary. The derivation of the resulting partition function is quite similar to that in the nonboundary cases. Now  Eq. \eqref{Z-BF-alg} can be denoted by
\begin{align}\label{BF-partition-function-Af-boundary}
 {\cal Z}^{\rm BF}(\Delta^*)
 &=\int{\rm d}g^+_{fv}{\rm d}g^-_{fv}\prod_{f\in\Delta^*}A^{\rm BF}_f(\{g^+_{fv},g^-_{fv};g^+_l,g^-_l\}).
\end{align} 
Integrating over the group elements $g_{ve}^\pm$ associated to the internal edges of $\Delta^*$ yields
\begin{align}
&A^{\rm BF}_f(\{g^+_{fv},g^-_{fv};g^+_l,g^-_l\})\notag\\
=&\makeSymbol{
\includegraphics[width=5.3cm]{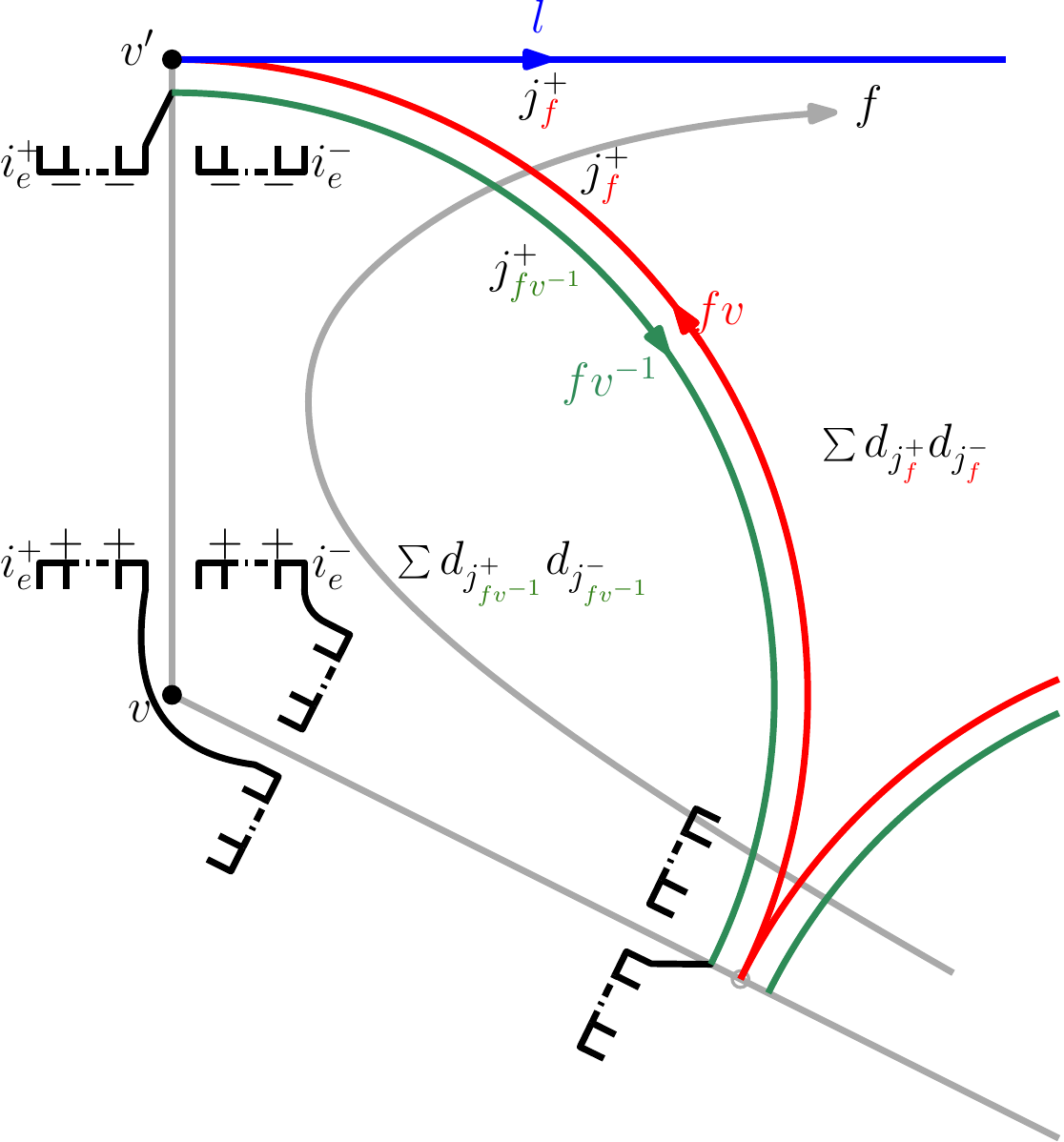}}.
\end{align}
Here the notations are the same as those for the nonboundary case.

Imposing the $Y$ map on the intertwiners associated to the vertices of $\gamma_v$ by Eq. \eqref{Y-map-intertwiner}, the partition function ${\cal Z}^{\rm BF}(\Delta^*)$ in Eq. \eqref{BF-partition-function-Af-boundary} can be promoted to the (Euclidean) generalized EPRL partition function
\begin{align}\label{Z-EPRL-A-boundary}
 {\cal Z}^{\rm EPRL}(\Delta^*):=\int{\rm d}g^+_{fv}{\rm d}g^-_{fv}\prod_{f\in\Delta^*}A^{\rm EPRL}_f(\{g^+_{fv},g^-_{fv};g^+_l,g^-_l\})
\end{align}
with
\begin{align}
 &A^{\rm EPRL}_f(\{g^+_{fv},g^-_{fv};g^+_l,g^-_l\}):=Y\cdot A^{\rm BF}_f(\{g^+_{fv},g^-_{fv};g^+_l,g^-_l\})\notag\\
 =&\makeSymbol{
\includegraphics[width=5.3cm]{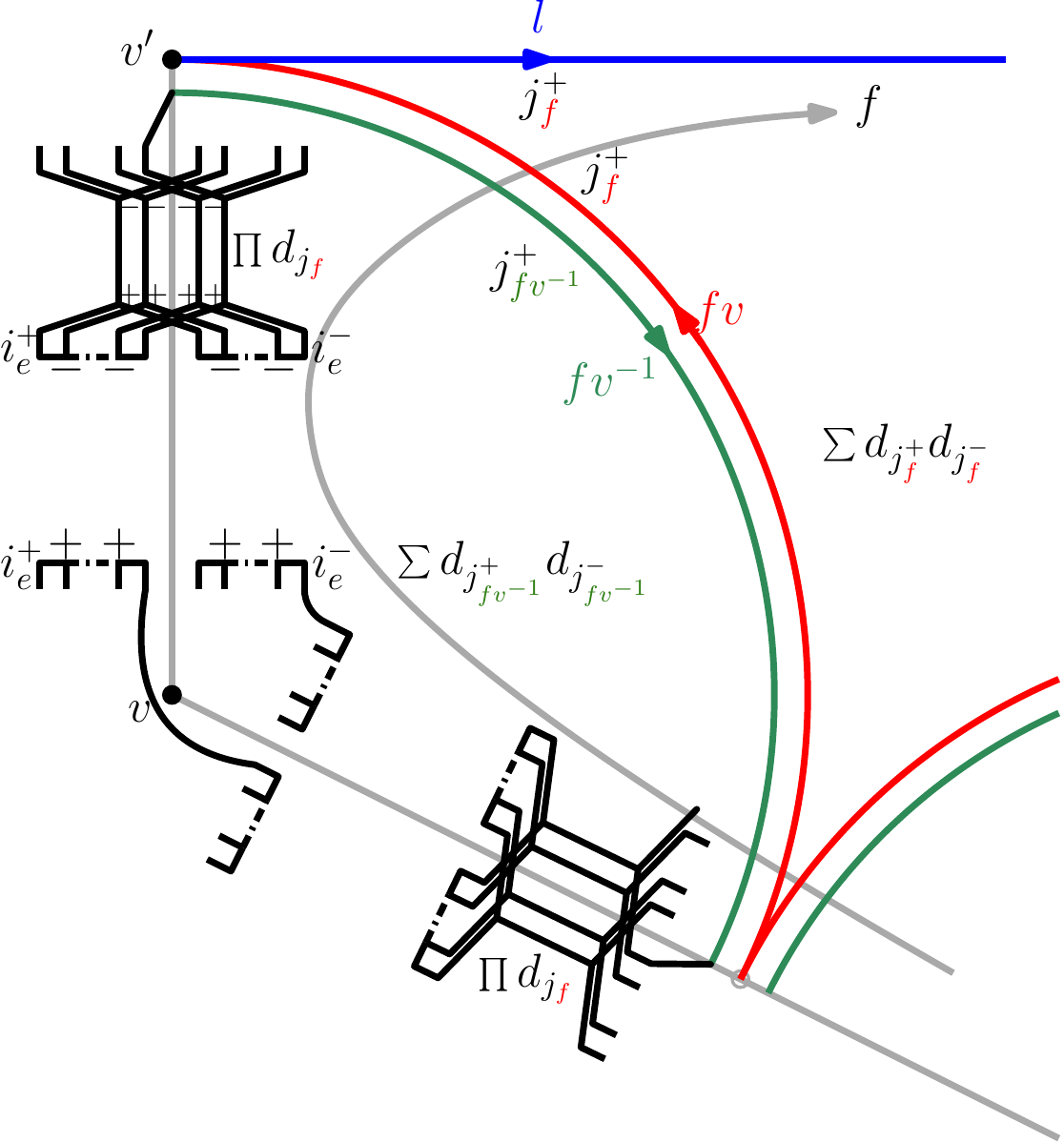}}\notag\\
=&\makeSymbol{
\includegraphics[width=5.3cm]{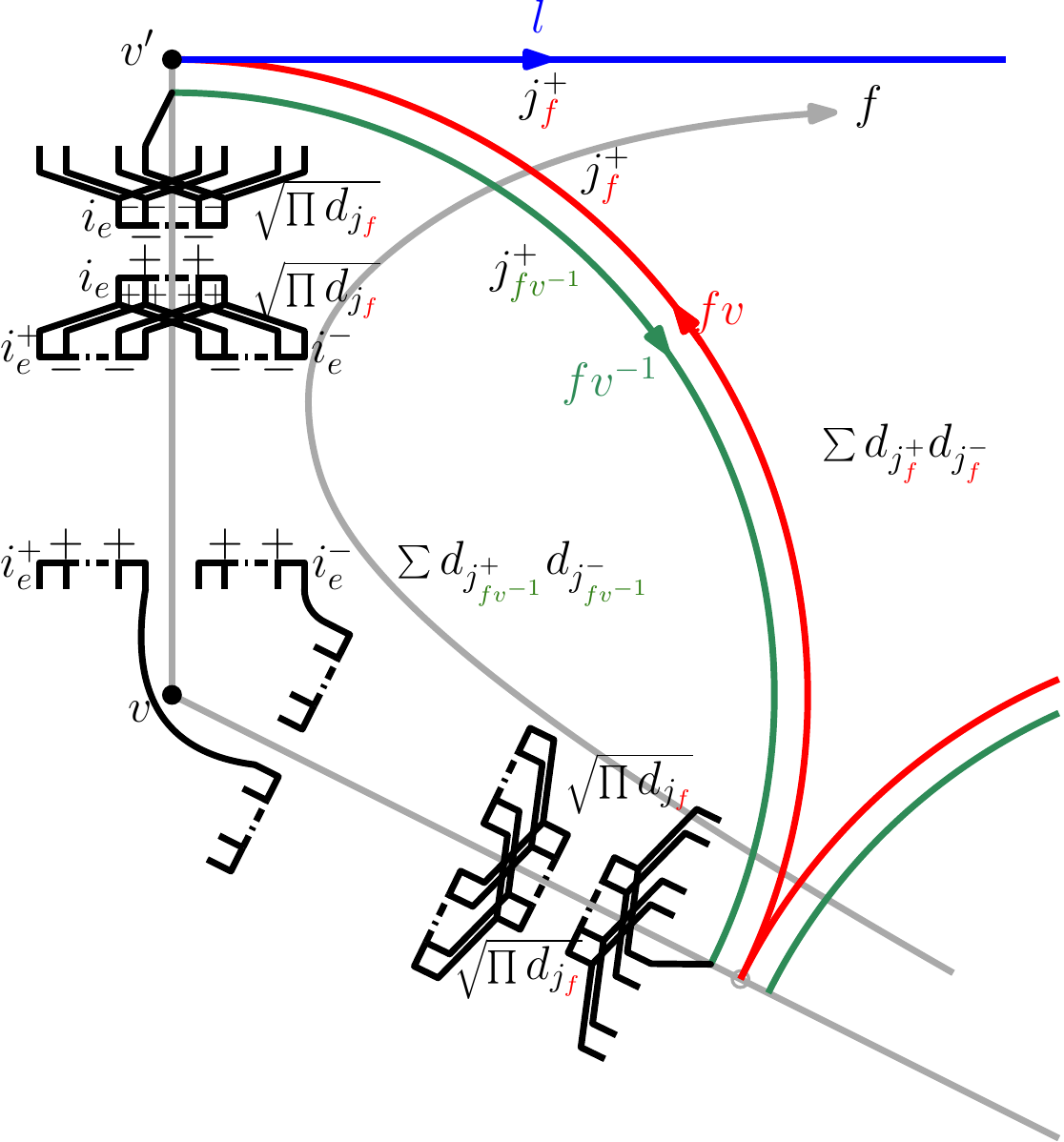}},
\end{align}
where we used Eq. \eqref{Y-map-intertwiner} in the second step, and Eq. \eqref{block-4} in the third step.

Using Eqs. \eqref{graph-two-edges-integral} and \eqref{circle-identity}, the integration over ${\rm d}g^+_{fv}{\rm d}g^+_{fv}$ in Eq. \eqref{Z-EPRL-A-boundary} yields
\begin{align}\label{intermediate-Z-boundary}
 {\cal Z}^{\rm EPRL}(\Delta^*)&=\prod_{f\in\Delta^*}\makeSymbol{
\includegraphics[width=5.3cm]{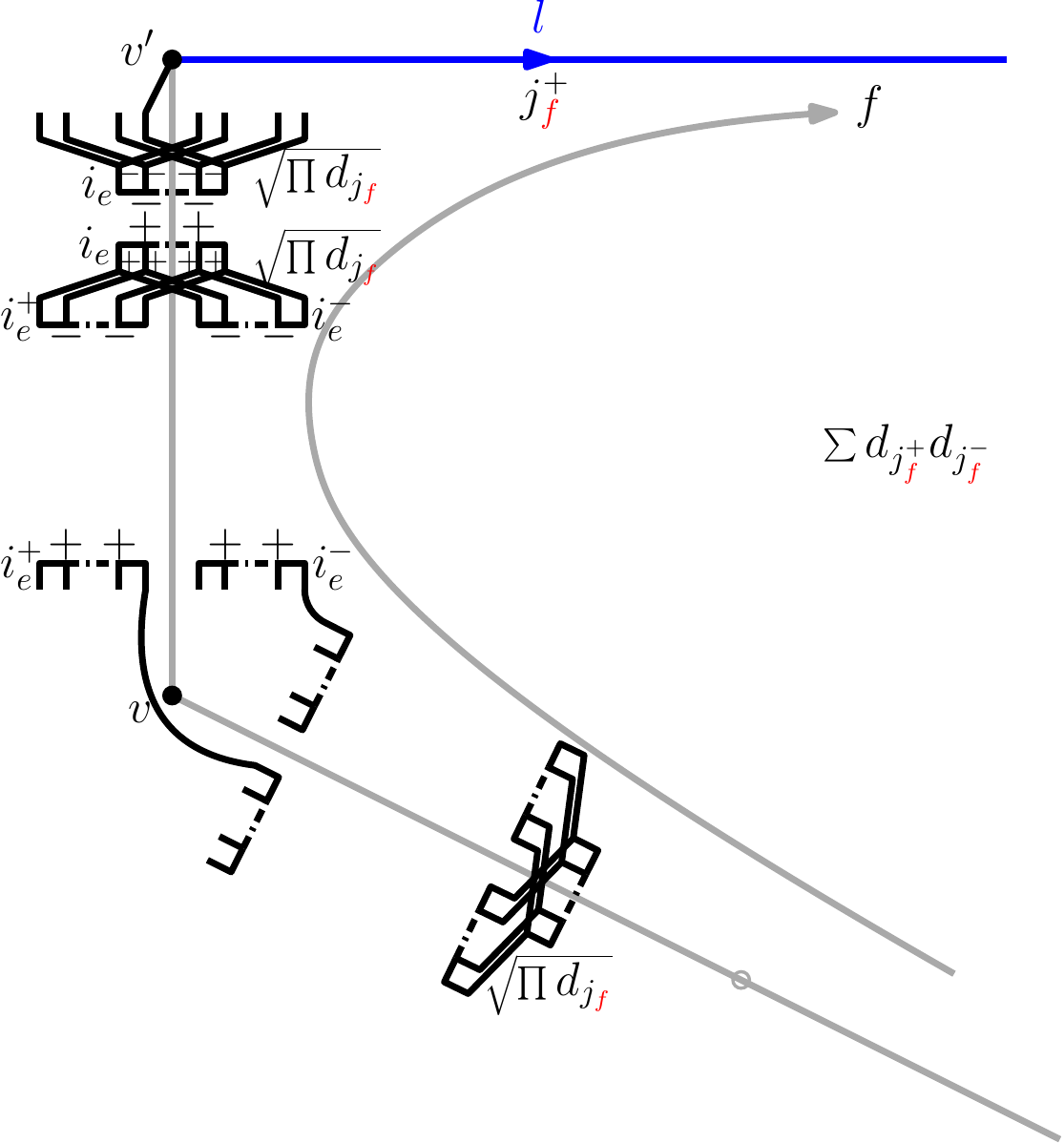}}.
\end{align}
Notice that the partition function \eqref{intermediate-Z-boundary} is a $SO(4)$ spin network
function induced on the boundaries. In order to match the $SU(2)$ spin network states of canonical LQG on the boundaries, one can restrict (or project) the $SO(4)$ group elements $(g^+_l,g^-_l)$ into $(g_l,g_l)$. After restricting the boundary elements and using Eqs.\eqref{holonomy-reps-couple} and \eqref{3j-orthogonality-2}, we have
\begin{align}\label{resulting-Z-boundary}
{\cal Z}^{\rm EPRL}(\Delta^*)&=\prod_{f\in\Delta^*}\makeSymbol{
\includegraphics[width=5.3cm]{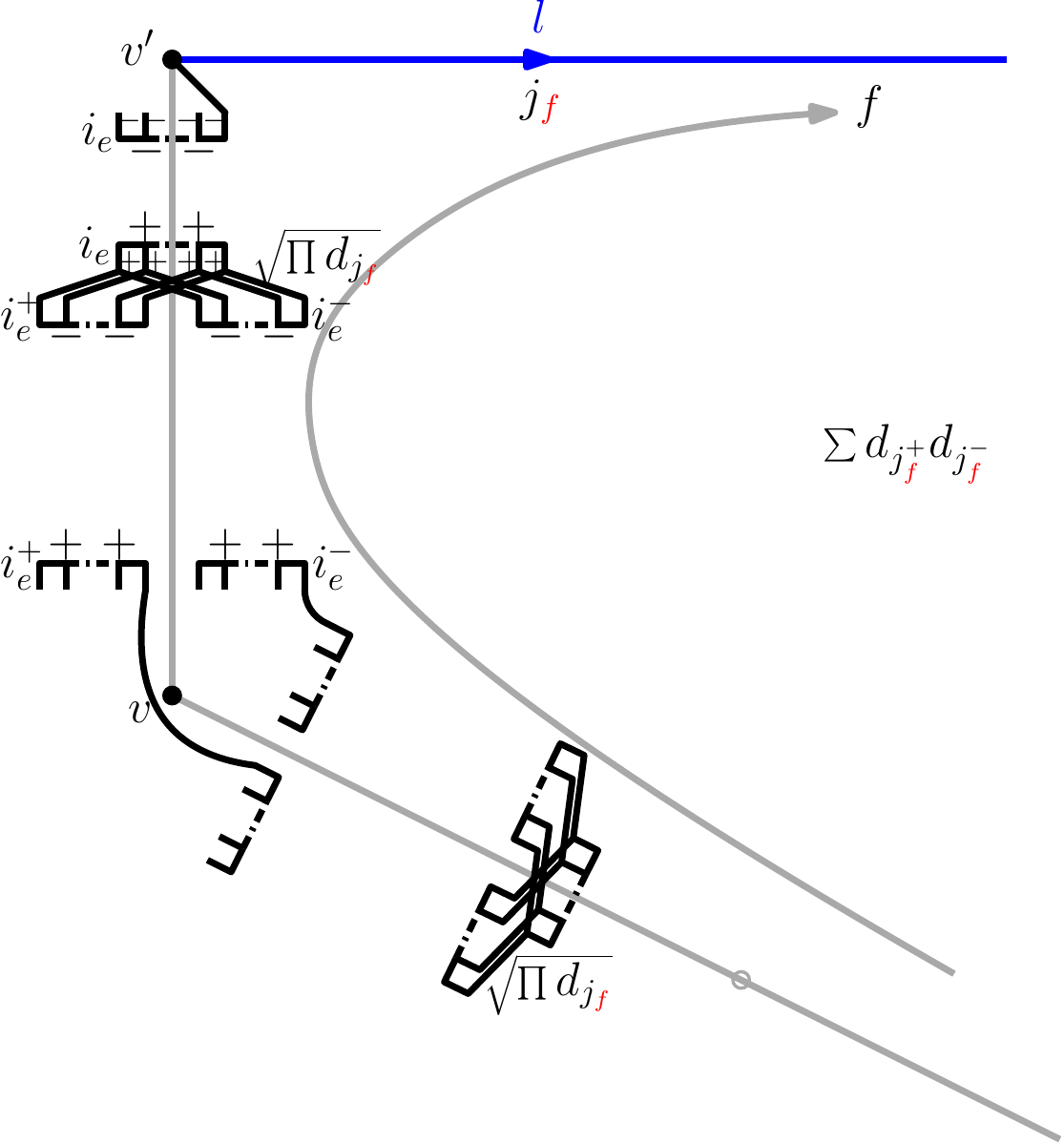}}.
\end{align}
It is worth noting that the intertwiners $i_e$ associated to the boundary vertices $v'\in\partial\Delta^*$ contracting with the representations $\pi_{j_l}(g_l)$ (the blue curved lines) of the boundary edges $l\in\partial\Delta^*$, multiplied by factors $\sqrt{d_{j_l}}$, form the normalized gauge-invariant spin network states on the boundary $\partial\Delta^*$. The resulting partition function ${\cal Z}^{\rm EPRL}(\Delta^*)$ expressed in the graphical formula \eqref{resulting-Z-boundary} can be uniquely transformed into the algebraic formula
\begin{align}\label{resulting-Z-boundary-algebraic}
 {\cal Z}^{\rm EPRL}(\Delta^*)&=\sum_{j^+_f,j^-_f}\prod_{f\in\Delta^*}d_{j^+_{f}}d_{j^-_{f}}\notag\\
 &\quad\times\prod_{v\in\Delta^*}\sum_{i^+_e,i^-_e,i_e}{\rm Tr}_v\left[\bigotimes_{e\in\partial v}(i^+_e\otimes i^-_e)\right]\prod_{e\in\partial v} f^{i_e}_{i^+_ei^-_e}\notag\\
 &\quad\times\sum_{j_l,i_v}\left(\prod_{l\in\partial \Delta^*}\frac{1}{\sqrt{d_{j_l}}}\right)T_{\partial\Delta^*,\vec{j}_l,\vec{i}_v}\left(\{g_l\}\right),
\end{align}
where $j_l:=j_f$, $i_v:=i_e$ denote the normalized gauge-invariant intertwiners associated to the boundary vertices $v\in\partial\Delta^*$, and $T_{\partial\Delta^*,\vec{j}_l,\vec{i}_v}\left(\{g_l\}\right)$ denote the normalized gauge-invariant spin network states on $\partial\Delta^*$. Given the ``in'' and ``out'' kinematical states $\psi_s$ and $\psi_{s'}$ on $\partial\Delta^*$, the transition amplitude between them is defined by
\begin{align}\label{boundary-amplitude}
&\sum_{\partial\Delta^*=\psi_s\cup\psi_{s'}}\langle \psi_{s'}|{\cal Z}^{\rm EPRL}(\Delta^*)|\psi_s\rangle\notag\\
:=&\sum_{\partial\Delta^*=\psi_s\cup\psi_{s'}}\sum_{j^+_f,j^-_f}\prod_{f\in\Delta^*}d_{j^+_{f}}d_{j^-_{f}}\prod_{v\in\Delta^*}\sum_{i^+_e,i^-_e,i_e}{\rm Tr}_v\left[\bigotimes_{e\in\partial v}(i^+_e\otimes i^-_e)\right]\prod_{e\in\partial v} f^{i_e}_{i^+_ei^-_e}\notag\\
 &\quad\times\left(\prod_{l\in\partial \Delta^*}\frac{1}{\sqrt{d_{j_l}}}\right),
\end{align}
where the summation $\sum_{\partial\Delta^*=\psi_s\cup\psi_{s'}}$ is taken over all possible $\Delta^*$ whose boundary states consist of $\psi_s$ and $\psi_{s'}$.

\section{Matrix elements of a Hamiltonian constraint operator}
\label{sec-III}
In the Hamiltonian formulation of GR, the Hamiltonian constraint for pure gravity in the Ashtekar-Barbero variables smeared with an arbitrary function $N$ on the spatial manifold $\Sigma$ reads \cite{Barbero:1994ap,Thiemann:2007pyv}
\begin{align}\label{eqn:full-hamilton}
H(N)&=\int_{\Sigma}{\rm d}^3x \frac{N\,\tilde{E}^a_i\tilde{E}^b_j}{2\kappa\sqrt{\textrm{det}(q)}}\left[{\epsilon^{ij}}_kF^{k}_{ab}
-2(-\zeta+\,\beta^2)K^i_{[a}K^j_{b]}\right],
\end{align}
where $\det(q)$ is the determinant of the spatial metric $q_{ab}$ on $\Sigma$, $\zeta$ denotes the spacetime signature such that $\zeta=-1$ and $\zeta=+1$ represent the Lorentzian and Euclidean cases respectively, $F^i_{ab}$ is the curvature of connection $A^i_a$, and $K^i_a$ represents the extrinsic curvature of $\Sigma$. In the Euclidean case of $\zeta=+1$, by taking the Immirzi parameter $\beta=1$, Eq. \eqref{eqn:full-hamilton} is reduced to the so-called Euclidean term
\begin{align}
H^{\rm E}(N)&:=\int_{\Sigma}{\rm d}^3x\, N\frac{\tilde{E}^a_i\tilde{E}^b_j}{2\kappa\sqrt{\textrm{det}(q)}}{\epsilon^{ij}}_kF^{k}_{ab}.
\end{align}
Different candidate Hamiltonian constraint operators corresponding to $H^{\rm E}(N)$ have been proposed for canonical LQG. Here we consider the Hamiltonian constraint operator $\hat{H}^{\rm E}(N)$ defined in \cite{Yang:2015zda}, which is well defined in certain partially diffeomorphism-invariant Hilbert space ${\cal H}_{\rm np4}$ and can be promoted as a symmetric operator. To simplify the discussion, we choose a corresponding regulated operator $\hat{H}^{\rm E}_\delta(N)$ in the kinematical Hilbert space ${\cal H}_{\rm kin}$ in following calculations. Since the operators $\hat{H}^{\rm E}_\delta(N)$ for different $\delta$ belong to the same diffeomorphism equivalent class \cite{Yang:2015zda}, all our calculations are also valid for the operator $\hat{H}^{\rm E}(N)$ in ${\cal H}_{\rm np4}$. By a special operator ordering, the action of $\hat{H}^{\rm E}_\delta(N)$ on a cylindrical function $f_\gamma$ over a graph $\gamma$ with edges outgoing from its vertices $v$ reads 
\begin{align}\label{eq:canonical-dynamics}
 \hat{H}^{\rm E}_\delta(N)\cdot f_\gamma&=-\frac{3(\beta\ell_{\rm p}^2)^2}{\kappa\chi(m)^2}\sum_{v\in V(\gamma)}N_v\,\hat{H}^{\rm E}_v\cdot f_\gamma,
\end{align}
where
\begin{align}
 \chi(x):=\sqrt{x(x+1)(2x+1)},
\end{align}
and
\begin{align}
\hat{H}^{\rm E}_v:=\sum_{e_i\cap e_j=v}\hat{H}^{\rm E}_{v,e_i,e_j}
\end{align}
with
\begin{align}
\hat{H}^{\rm E}_{v,e_i,e_j}&:=\epsilon_{kls}{\rm Tr}_m(\tau_kg_{\alpha_{ij}})J^l_iJ^s_j\widehat{V^{-1}}_v\notag\\
&=-{\rm i}\epsilon_{\mu\nu\rho}{\rm Tr}_m(\tau_\mu g_{\alpha_{ij}})J^\nu_iJ^\rho_j\widehat{V^{-1}}_v\notag\\
&=-{\rm i}\epsilon_{\mu\nu\rho}{[\pi_m(\tau_\mu)]^A}_B {[\pi_m(g_{\alpha_{ij}})]^B}_AJ^\nu_iJ^\rho_j\widehat{V^{-1}}_v.
\end{align}
Here $g_{\alpha_{ij}}$ represents the hononomies along loops $\alpha_{ij}=e_i^1\circ a_{ij}\circ \left(e_j^1\right)^{-1}$ based at the vertices $v$ consisting of two segments $e^1_i$ and $e^1_j$ of $e_i$ and $e_j$, such that $e_i=e^1_i\circ e^2_i$ and $e_j=e^1_j\circ e^2_j$, the arc $a_{ij}$ connects the two endpoints of $e^1_i$ and $e^1_j$, the orientation of $\alpha_{ij}$ has been specially chosen such that it agrees with the one induced by $\Sigma$, $J^l_i\equiv J^l_{e_i}$ is the self-adjoint right-invariant vector field on a copy of $SU(2)$ associated to $e_i$, $\widehat{V^{-1}}_v$ denotes the inverse volume operator
\begin{align}\label{eq:inverse-volume}
\widehat{V^{-1}}:=\lim_{\lambda\rightarrow0}\frac{\hat{V}}{\hat{V}^2+(\lambda\ell_{\rm p}^3)^2}
\end{align}
acting at vertices $v$, $\epsilon_{\mu\nu\rho}$ ($\mu,\nu,\rho=0,+1,-1$) is the Levi-Civita symbol defined by $\epsilon_{-1
\;0\;+1}=1$, and $\tau_\mu$ $(\mu=0,\pm1)$ is the spherical tensor, which is related to the basis $\tau_i:=-{\rm i}\sigma_i/2$ $(i=1,2,3)$ of $su(2)$ with $\sigma_i$ being the Pauli matrices by
\begin{align}
\tau_0:=\tau_3,\qquad \tau_{\pm}:=\mp\frac{1}{\sqrt{2}}(\tau_1\pm{\rm i}\tau_2).
\end{align}
Intuitively, the operator $\hat{H}^{\rm E}_v$ acts on $\gamma$ by attaching an arc $a_{ij}$ to each pair $(e_i, e_j)$ of edges such that the resulting loop $\alpha_{ij}$ has a positive orientation.  Notice that the operator $\widehat{V^{-1}}_v$ in Eq. \eqref{eq:inverse-volume} vanishes at a gauge-invariant vertex $v$ with valence less than four by the property of the volume operator $\hat{V}$ \cite{Ashtekar:1997fb}. Consider a simply graph $\gamma$ with a noncoplanar vertex $v$ and four edges $e_1,\cdots,e_4$ starting from $v$. To specify a spin network state to $\gamma$, one needs to specify spins $j_1,\cdots,j_4$ to its edges $e_1,\cdots,e_4$ and a gauge-invariant intertwiner to $v$. To specify the intertwiner, one needs to choose a coupling scheme for $j_1,\cdots,j_4$ and an intermediate coupling spin. Different coupling schemes are related by the $6j$-symbol. For example, the following formula relates the two different coupling schemes for the intertwiners associated to $v$ \cite{Brink:1968bk,Yang:2015wka}
\begin{align}
\makeSymbol{
\includegraphics[width=2cm]{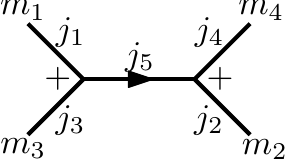}}=&\sum_{j_6}d_{j_6}(-1)^{j_1-j_2+j_3+j_4}
\begin{Bmatrix}
j_1 & j_4 & j_6\\
j_2 & j_3 & j_5
\end{Bmatrix}\notag\\
 &\qquad\times\makeSymbol{
\includegraphics[width=2cm]{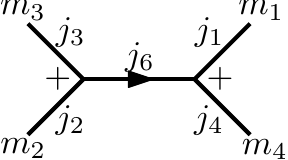}}\,.
\end{align}
Given a gauge-invariant spin network state $\psi_s$ on $\gamma$ with a specified intertwiner at $v$, the action of $\hat{H}^{\rm E}_v$ on $\psi_s$ is given by
\begin{align}\label{H-action-arguement}
&\hat{H}^{\rm E}_v\cdot\psi_s=\sum_{e_i\cap e_j=v}\hat{H}^{\rm E}_{v,e_i,e_j}\cdot\psi_s\notag\\
=&\sum_{e_i\cap e_j=v}\hat{H}^{\rm E}_{v,e_i,e_j}\cdot\sum_{i}f^i_s\sqrt{d_{j_i}d_{j_j}d_{j_k}d_{j_l}d_i}\makeSymbol{
\includegraphics[height=1.8cm]{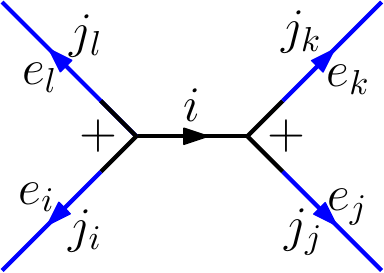}}\notag\\
=&\sum_{i}f^i_s\sum_{e_i\cap e_j=v}\hat{H}^{\rm E}_{v,e_i,e_j}\cdot\sqrt{d_{j_i}d_{j_j}d_{j_k}d_{j_l}d_i}\makeSymbol{
\includegraphics[height=1.8cm]{figures/Hamiltonian/YM-0}},
\end{align}
where $f^i_s$ are the expanding factors involving $6j$-symbols. Hence, one only needs to evaluate the action of $\hat{H}^{\rm E}_{v,e_i,e_j}$ on the normalized spin network state
\begin{align}\label{psi-i}
\psi_i=\sqrt{d_{j_i}d_{j_j}d_{j_k}d_{j_l}d_i}\makeSymbol{
\includegraphics[height=1.8cm]{figures/Hamiltonian/YM-0}}.
\end{align}
Notice that the Levi-Civita symbol $\epsilon_{\mu\nu\rho}$ is related to the $3j$-symbols by \cite{Yang:2015wka}
\begin{align}
 \epsilon_{\mu\nu\rho}=\chi(1)
\begin{pmatrix}
1 & 1 & 1\\
\mu & \nu & \rho
\end{pmatrix}=\chi(1)\makeSymbol{
\includegraphics[height=1cm]{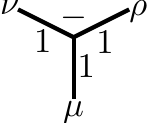}},
\end{align}
the spherical tensors $\tau_\mu$ can be represented by \cite{Yang:2015wka}
\begin{align}\label{spher-rep-graph}
{[\pi_j(\tau_\mu)]^A}_B&={\rm i}\chi(j)\makeSymbol{
\includegraphics[width=1.7cm]{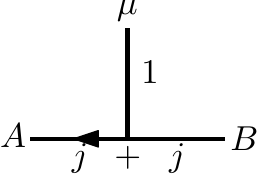}}={\rm i}\chi(j)\makeSymbol{
\includegraphics[width=1.7cm]{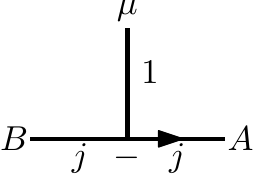}}\,,
\end{align}
and the action of $J^\mu_i$ on a spin network state is determined by its action on the corresponding intertwiner as \cite{Yang:2015wka}
\begin{align}
&J^\mu_i \prod_{i=2}^{n-2}\sqrt{d_{a_i}}\makeSymbol{
\includegraphics[width=4.7cm]{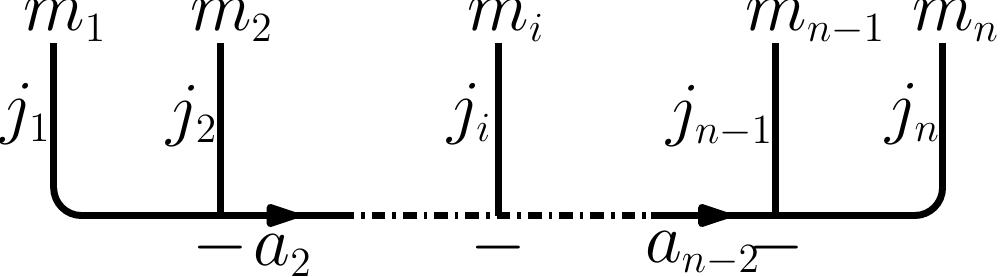}}\notag\\
=&\chi(j_i)\prod_{i=2}^{n-2}\sqrt{d_{a_i}}\makeSymbol{
\includegraphics[width=4.7cm]{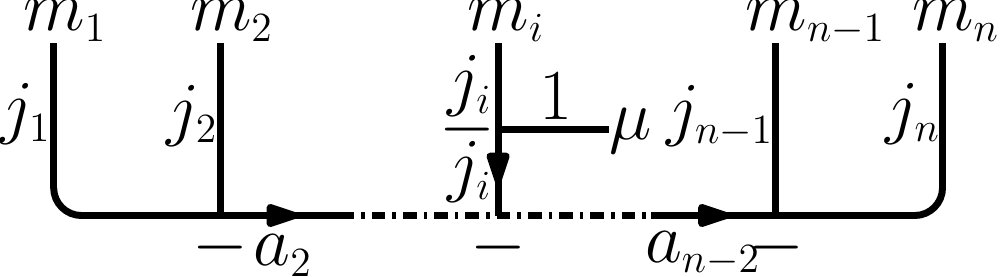}}.
\end{align}
Thus, the action of $\hat{H}^{\rm E}_{v,e_i,e_j}$ on $\psi_i$ can be calculated as
\begin{align}\label{YM-Hamitonian-action}
&\hat{H}^{\rm E}_{v,e_i,e_j}\sqrt{d_i}\makeSymbol{
\includegraphics[height=1.8cm]{figures/Hamiltonian/YM-0}}\notag\\
=&-{\rm i}\epsilon_{\mu\nu\rho}{[\pi_m(\tau_\mu)]^A}_B {[\pi_m(g_{\alpha_{ij}})]^B}_AJ^\nu_iJ^\rho_j\sum_k\left(\widehat{V^{-1}}_v\right)_i^k\notag\\
&\hspace{1cm}\times\sqrt{d_k}\makeSymbol{
\includegraphics[height=1.8cm]{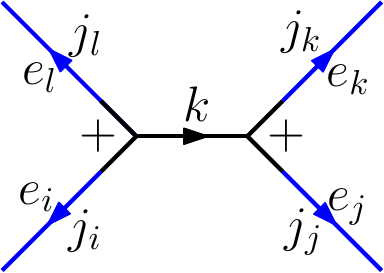}}\notag\\
=&\chi(1)\chi(m)\chi(j_i)\chi(j_j)\sum_k\left(\widehat{V^{-1}}_v\right)_i^k\notag\\
&\hspace{1cm}\times\sqrt{d_k}\sum_{a,b}d_ad_b\makeSymbol{
\includegraphics[height=4.4cm]{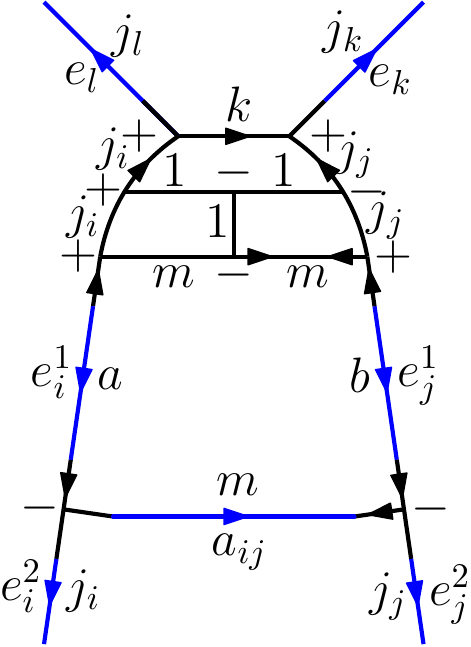}}\notag\\
=&\chi(1)\chi(m)\chi(j_i)\chi(j_j)\sum_k\left(\widehat{V^{-1}}_v\right)_i^k\sqrt{d_k}\sum_{a,b}d_ad_b\notag\\
&\times\sum_cd_c(-1)^{b+c+j_j-1}\begin{Bmatrix}
c & j_i & a\\
j_i & m & 1
\end{Bmatrix}\begin{Bmatrix}
c & j_j & b\\
j_j & m & 1
\end{Bmatrix}
\begin{Bmatrix}
c & m & 1\\
1 & 1 & m
\end{Bmatrix}\notag\\
&\times\sum_t d_t(-1)^{j_l-j_i-k}(-1)^{t+b-j_k}\begin{Bmatrix}
t & j_l & a\\
j_i & c & k
\end{Bmatrix}
\begin{Bmatrix}
t & j_k & b\\
j_j & c & k
\end{Bmatrix}\notag\\
&\qquad\times\makeSymbol{
\includegraphics[height=4cm]{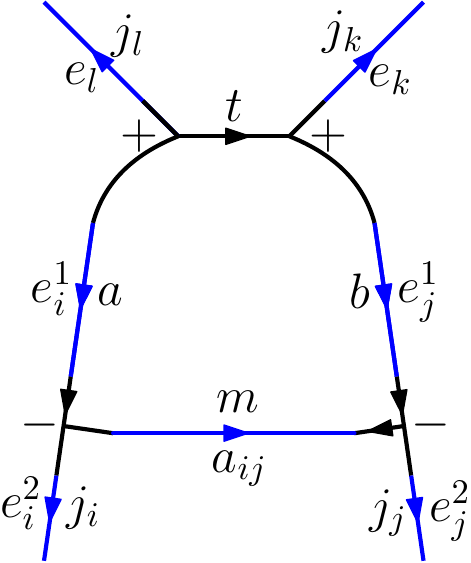}},
\end{align}
where $\left(\widehat{V^{-1}}_v\right)_i^k\equiv\langle i^k_v|\widehat{V^{-1}}_v|i^i_v\rangle$ denote the matrix elements of $\widehat{V^{-1}}_v$ between intertwiners $i^i_v$ and $i^k_v$ associated to $v$ and labeled by the intermediate angular momenta $i$ and $k$, and in the last step we used
\begin{align}\label{eq:action-Hamiltonian-simplify}
&\makeSymbol{
\includegraphics[height=1.9cm]{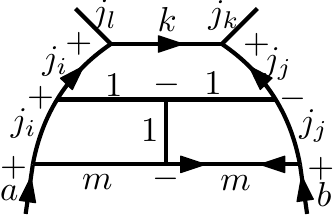}}=\sum_{c,d}d_cd_d\makeSymbol{
\includegraphics[height=1.85cm]{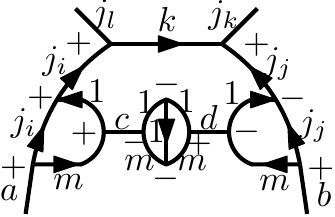}}\notag\\
=&\sum_cd_c(-1)^{b+c+j_j-1}\begin{Bmatrix}
c & j_i & a\\
j_i & m & 1
\end{Bmatrix}\begin{Bmatrix}
c & j_j & b\\
j_j & m & 1
\end{Bmatrix}
\begin{Bmatrix}
c & m & 1\\
1 & 1 & m
\end{Bmatrix}\notag\\
&\hspace{0.5cm}\times\makeSymbol{
\includegraphics[height=1.85cm]{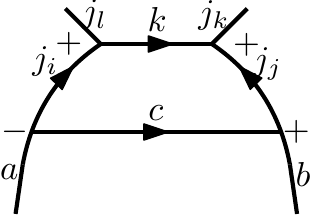}}\notag\\
=&\sum_cd_c(-1)^{b+c+j_j-1}\begin{Bmatrix}
c & j_i & a\\
j_i & m & 1
\end{Bmatrix}\begin{Bmatrix}
c & j_j & b\\
j_j & m & 1
\end{Bmatrix}
\begin{Bmatrix}
c & m & 1\\
1 & 1 & m
\end{Bmatrix}\notag\\
&\hspace{0.5cm}\times\sum_t d_t\makeSymbol{
\includegraphics[height=1.85cm]{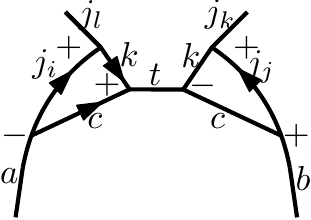}}\notag\\
=&\sum_cd_c(-1)^{b+c+j_j-1}\begin{Bmatrix}
c & j_i & a\\
j_i & m & 1
\end{Bmatrix}\begin{Bmatrix}
c & j_j & b\\
j_j & m & 1
\end{Bmatrix}
\begin{Bmatrix}
c & m & 1\\
1 & 1 & m
\end{Bmatrix}\notag\\
&\hspace{0.5cm}\times\sum_t d_t(-1)^{j_l-j_i-k}(-1)^{t+b-j_k}
\begin{Bmatrix}
t & j_l & a\\
j_i & c & k
\end{Bmatrix}
\begin{Bmatrix}
t & j_k & b\\
j_j & c & k
\end{Bmatrix}\notag\\
&\hspace{1.4cm}\times\makeSymbol{
\includegraphics[height=1.2cm]{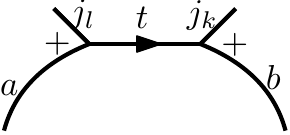}}.
\end{align}
Note that to derive Eq. \eqref{eq:action-Hamiltonian-simplify}, we used Eqs. \eqref{3j-orthogonality-1}, \eqref{three-arrow-adding}, and \eqref{two-arrow-cancel} in the first step, the identities
\begin{align}
\makeSymbol{
\includegraphics[height=1.5cm]{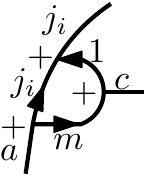}}&=(-1)^{m+1+c}(-1)^{2j_i+1}(-1)^{2j_i}(-1)^{m+a+j_i}\makeSymbol{
\includegraphics[height=1.7cm]{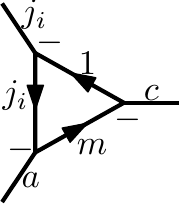}}\notag\\
&=(-1)^{2m+j_i+a+c}
\begin{Bmatrix}
c & j_i & a\\
j_i & m & 1
\end{Bmatrix}
\makeSymbol{
\includegraphics[height=0.95cm]{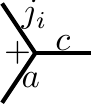}}\notag\\
&=(-1)^{2m}\begin{Bmatrix}
c & j_i & a\\
j_i & m & 1
\end{Bmatrix}
\makeSymbol{
\includegraphics[height=1.15cm]{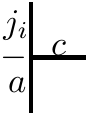}},\\
\makeSymbol{
\includegraphics[height=1.5cm]{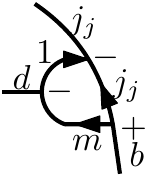}}&=(-1)^{j_j+m+b}(-1)^2(-1)^{2m}
\makeSymbol{
\includegraphics[height=1.7cm]{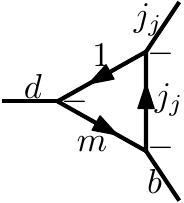}}\notag\\
&=(-1)^{b+j_j-m}\begin{Bmatrix}
d & j_j & b\\
j_j & m & 1
\end{Bmatrix}\makeSymbol{
\includegraphics[height=1.1cm]{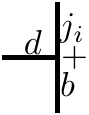}},\\
\makeSymbol{\includegraphics[height=1.1cm]{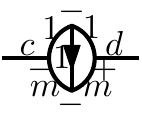}}&=\frac{\delta_{c,d}}{d_d}\makeSymbol{\includegraphics[height=1.1cm]{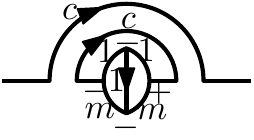}}\notag\\
&= \frac{\delta_{c,d}}{d_d}(-1)^{m+1-c}\makeSymbol{\includegraphics[height=1.4cm]{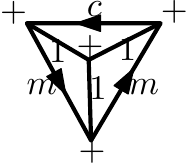}}\makeSymbol{\includegraphics[height=0.3cm]{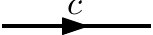}}\notag\\
&= \frac{\delta_{c,d}}{d_d}(-1)^{m+1-c}\begin{Bmatrix}
c & m & 1\\
1 & 1 & m
\end{Bmatrix}
\makeSymbol{\includegraphics[height=0.3cm]{figures/Hamiltonian/YM-id-step-II-3-4}},
\end{align}
in the second step, Eq. \eqref{3j-orthogonality-1} in the third step, and Eqs. \eqref{3j-orientation-change-graph}, \eqref{three-arrow-adding}--\eqref{two-arrow-result}, and \eqref{6j-def-graph} in the last step. Equation \eqref{YM-Hamitonian-action} shows that the action of $\hat{H}^{\rm E}_{v,e_i,e_j}$ on $\psi_i$ can be linearly expanded in terms of the normalized spin network states
\begin{align}\label{eq:phi-alpha}
\psi_t&=\sqrt{d_{j_i}d_{j_j}d_{j_k}d_{j_l}d_ad_bd_md_t}\makeSymbol{
\includegraphics[height=4cm]{figures/Hamiltonian/YM-3}}.
\end{align}
In the definition of $\hat{H}^{\rm E}_\delta(N)$ in \cite{Yang:2015zda}, the spin $m$ of arc $a_{ij}$ was chosen in such a way that neither the spin $a$ of $e^1_i$ nor the spin $b$ of $e^1_j$ vanishes. Then the matrix elements $\langle \psi_t|\hat{H}^{\rm E}_{v,e_i,e_j}|\psi_i\rangle$ can be easily calculated as
\begin{align}\label{H-matrix-element}
&\langle \psi_t|\hat{H}^{\rm E}_{v,e_i,e_j}|\psi_i\rangle\notag\\
=&\chi(1)\chi(m)\chi(j_i)\chi(j_j)\sum_k\sqrt{d_k}\left(\widehat{V^{-1}}_v\right)_i^k d_ad_b\notag\\
&\times\sum_cd_c(-1)^{b+c+j_j-1}
\begin{Bmatrix}
c & j_i & a\\
j_i & m & 1
\end{Bmatrix}
\begin{Bmatrix}
c & j_j & b\\
j_j & m & 1
\end{Bmatrix}
\begin{Bmatrix}
c & m & 1\\
1 & 1 & m
\end{Bmatrix} 
 \notag\\
&\times d_t(-1)^{j_l-j_i-k}(-1)^{t+b-j_k}
\begin{Bmatrix}
t & j_l & a\\
j_i & c & k
\end{Bmatrix}
\begin{Bmatrix}
t & j_k & b\\
j_j & c & k
\end{Bmatrix}\notag\\
&\times\frac{1}{\sqrt{d_ad_bd_md_t}}.
\end{align}

\section{Relation in quantum dynamics}
\label{sec-IV}
The quantum dynamics in covariant LQG is encoded in the partition function for a given SFM, while the quantum dynamics in canonical LQG is determined by the Hamiltonian constraint operator obtained from a suitable quantization procedure. To check the consistency of the two formulations of quantum dynamics is a crucial task. On one hand, in canonical LQG, one expects to construct the physical inner product in the physical Hilbert space by an antilinear rigging map \cite{Thiemann:2007pyv}
\begin{align}
\eta: {\cal D}_{\rm kin}&\rightarrow {\cal D}^*_{\rm phys};\quad
 \psi_s\mapsto \eta(\psi_s),
\end{align}
where ${\cal D}_{\rm kin}$ represents a certain dense domain of ${\cal H}_{\rm kin}$, and the space ${\cal D}^*_{\rm phys}$ of solutions to the quantum Hamiltonian constraint is regarded as a subspace of the algebraic dual ${\cal D}^*_{\rm kin}$ of ${\cal D}_{\rm kin}$. Thus, the physical inner product can be defined as
\begin{align}
\langle\eta(\psi_s)|\eta(\psi_{s'})\rangle_{\rm phys}&:=\left[\eta(\psi_{s'})\right](\psi_s).
\end{align}
On the other hand, a SFM can naturally provide a rigging map by the transition amplitudes as
\begin{align}
\langle\eta(\psi_s)|\eta(\psi_{s'})\rangle_{\rm phys}&:=\sum_{\partial\Delta^*=\psi_s\cup\psi_{s'}}\langle\psi_{s'}|{\cal Z}^{\rm SFM}(\Delta^*)|\psi_s\rangle.
\end{align}
Hence the rigging map can be defined by
\begin{align}\label{rigging-map}
\eta(\psi_s):=\sum_{\psi_t\in{\cal D}_{\rm kin}}\sum_{\partial\Delta^*=\psi_s\cup\psi_t}\langle\psi_s|{\cal Z}^{\rm SFM}(\Delta^*)|\psi_t\rangle\langle\psi_t|,
\end{align}
for all $\psi_s\in{\cal D}_{\rm kin}$. From the viewpoint of canonical LQG, the physical inner product should satisfy
\begin{align}\label{quantum-dynamical-relation}
&\langle\hat{H}'\eta(\psi_s)|\eta(\psi_{s'})\rangle_{\rm phys}=\left[\eta(\psi_{s'})\right](\hat{H}\psi_s)\notag\\
=&\sum_{\psi_t\in{\cal D}_{\rm kin}}\sum_{\partial\Delta^*=\psi_t\cup\psi_{s'}}\langle\psi_{s'}|{\cal Z}^{\rm SFM}(\Delta^*)|\psi_t\rangle\langle\psi_t|\hat{H}|\psi_s\rangle\notag\\
=&0,
\end{align}
for all $\psi_s,\psi_{s'}\in{\cal D}_{\rm kin}$, where $\hat{H}'$ defined on ${\cal D}^*_{\rm kin}$ is the dual of the Hamiltonian constraint operator $\hat{H}$ defined on ${\cal D}_{\rm kin}$. Eq. \eqref{quantum-dynamical-relation} implies the consistency between the covariant and canonical formulations of quantum dynamics in the sense that the Hamiltonian constraint of the latter is weakly satisfied for the physical states of the former.

We now check whether such a consistency exists between the covariant dynamics determined by Eq. \eqref{resulting-Z-boundary} and the canonical dynamics given by Eq. \eqref{eq:canonical-dynamics}. Let us consider the simple case in which $\Delta^*$ has only one internal vertex $v$, and focus on the Euclidean sector with the Immirzi parameter $\beta=1$. It is easy to see that Eq. \eqref{quantum-dynamical-relation} is automatically satisfied for $\psi_s$ on a graph with valence less than four, due to the property of $\widehat{V^{-1}}$ appeared in $\hat{H}^{\rm E}$. To check the consistency for the nontrivial cases, we focus on a graph with vertices of valence more than three. To make thing simple from which one can find crucial ingredient for the identity \eqref{quantum-dynamical-relation}, we first consider a graph $\gamma$ with a 4-valent vertex, and assume that $\psi_s$ and $\psi_{s'}$ are tightly related as shown below. We will go back to the multivalent cases and relax the restraint on $\psi_s$ and $\psi_{s'}$ at the end of this section. Let $\psi_s$ be a normalized gauge-invariant spin network state on a graph $\gamma$ consisting of one vertex $v_1$ and four edges $e_1,\cdots,e_4$ starting from $v_1$, and $\psi_{s'}$ be another spin network state on a graph $\gamma'$ consisting of one vertex $v'_1$ and four edges $e_1^{-1},\cdots,e_4^{-1}$ incoming to $v'_1$. Both of $e_i$ and $e_i^{-1}$ are labeled by the same spin $j_i$. The intertwiner $i_{v'_1}$ associated to $v'_1$ can be different from the intertwiner $i_{v_1}$ associated to $v_1$. By Eq. \eqref{H-action-arguement}, Eq. \eqref{quantum-dynamical-relation} can be expressed as
\begin{align}\label{quantum-dynamical-relation-reduce}
&\sum_{i,i'}f^i_sf^{i'}_{s'}\sum_{e_i\cap e_j=v_1}\sum_{\partial\Delta^*=\psi_t\cup\psi_{i'}}\sum_{\psi_t}\langle\psi_{i'}|{\cal Z}^{\rm EPRL}(\Delta^*)|\psi_t\rangle\langle\psi_t|\hat{H}^{\rm E}_{v_1,e_i,e_j}|\psi_i\rangle\notag\\
&=0,
\end{align}
where $\psi_{i'}$ is the normalized gauge-invariant spin network state with the same coupling scheme as that of $\psi_i$, but possible different intermediate coupling spins from that of $\psi_i$. It should be noted that only those $\psi_t$ given by Eq. \eqref{eq:phi-alpha} have nontrivial contribution to the matrix elements of $\hat{H}^{\rm E}_{v_1,e_i,e_j}$ in \eqref{quantum-dynamical-relation-reduce}. Note also that the partition function in Eq. \eqref{quantum-dynamical-relation-reduce} is defined on the $\Delta^*$ with some boundary states $\psi_{i'}\cup\psi_t$ and only one internal vertex $v$. The quantum dynamics can be presented by a visual picture as
\begin{align}
 \makeSymbol{
\includegraphics[width=0.45\textwidth]{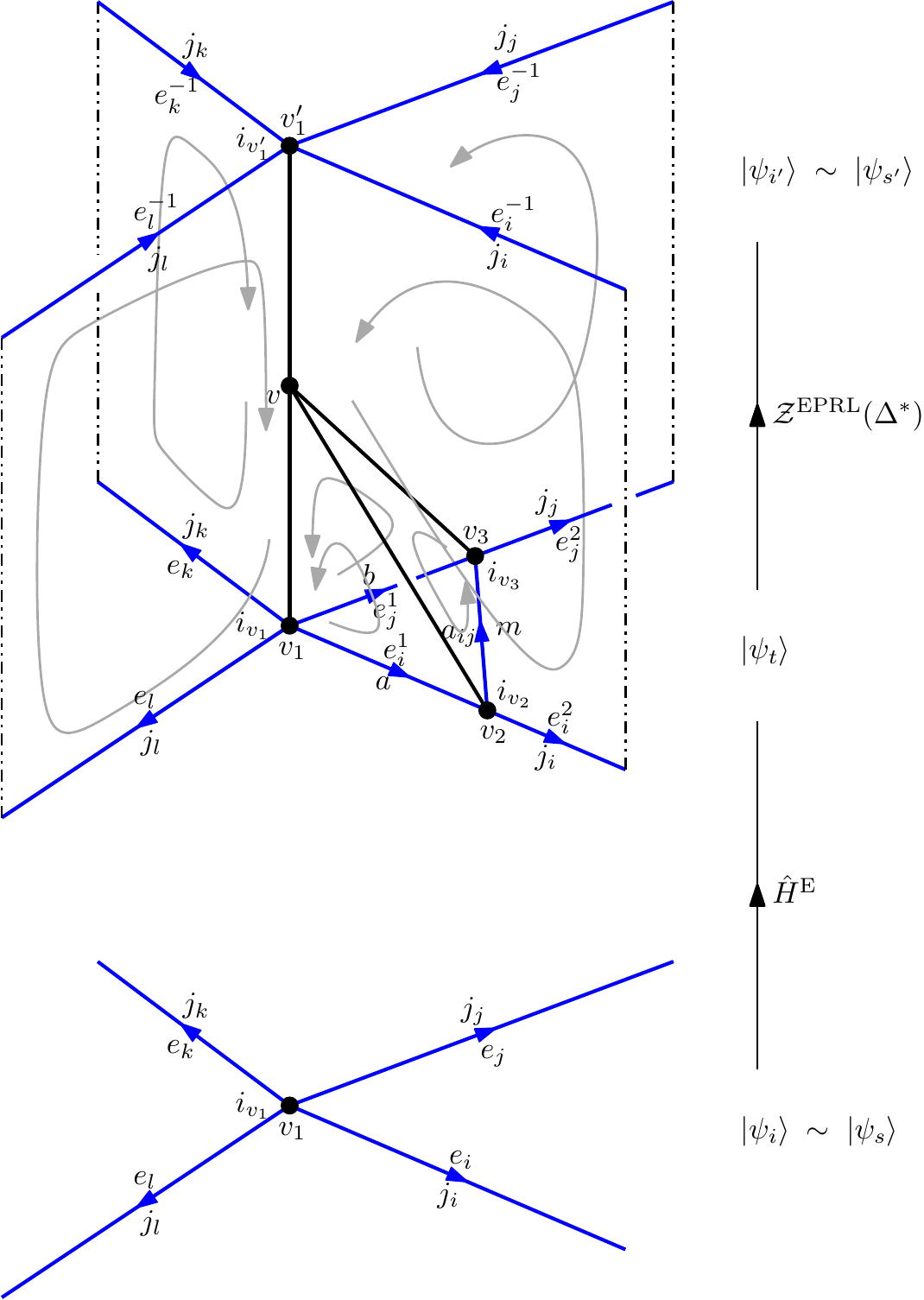}}\notag .
\end{align}
To see whether Eq. \eqref{quantum-dynamical-relation-reduce} is satisfied, it is sufficient to check whether one has equation
\begin{align}\label{consistency-check}
\sum_{\psi_t}\langle\psi_{i'}|{\cal Z}^{\rm EPRL}(\Delta^*)|\psi_t\rangle\langle\psi_t|\hat{H}^{\rm E}_{v_1,e_i,e_j}|\psi_i\rangle=0.
\end{align}

Let us first compute the transition amplitude $\langle\psi_{i'}|{\cal Z}^{\rm EPRL}(\Delta^*)|\psi_t\rangle$. By Eq. \eqref{eq:simplicity-beta-1}, the condition $\beta=1$ implies 
\begin{align}
 j^+_f&=j_f=j_l,\quad j^-_f=0, \quad \forall l\in\partial \Delta^*\cap f.
\end{align}
This condition simplifies the fusion functions $f^{i_e}_{i^+_ei^-_e}$ as well as the vertex amplitudes in the transition amplitude \eqref{boundary-amplitude}. For example, the fusion function associated to the internal edge linking $v$ and $v_1$ reads
\begin{align}
\makeSymbol{
\includegraphics[height=1.9cm]{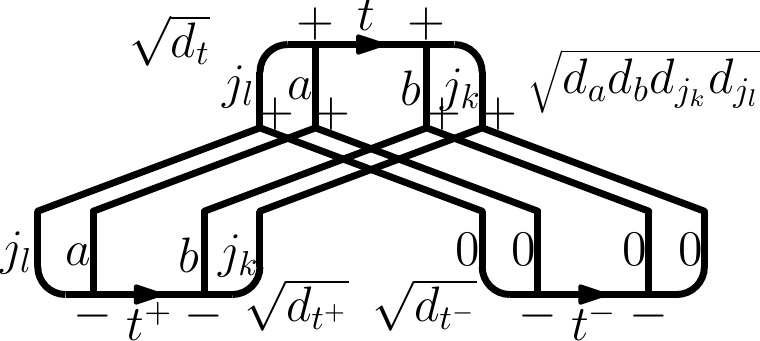}}&=\delta_{t^-,0}\makeSymbol{
\includegraphics[height=1.9cm]{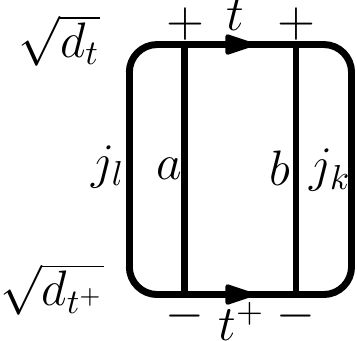}}\notag\\
&=\delta_{t^+,t}\delta_{t^-,0},
\end{align}
where Eqs. \eqref{arrow-3j}, \eqref{two-arrow-cancel}, \eqref{3j-orthogonality-2}, and \eqref{circle-rule} were used. The vertex amplitude $A_v$ associated to the internal vertex $v$ is reduced to
\begin{align}\label{A-v-graph-simplify}
A_v&={\rm Tr}(i_{v_1}\otimes i_{v_2}\otimes i_{v_3}\otimes i_{v'_1})=\makeSymbol{
\includegraphics[height=4.8cm]{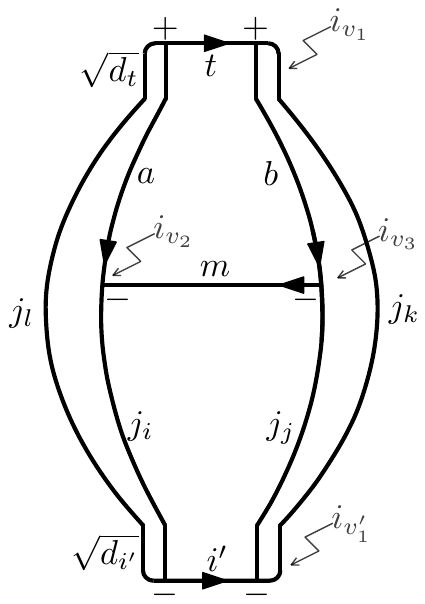}}\notag\\
&=\sqrt{d_t d_{i'}}\makeSymbol{
\includegraphics[height=1.6cm]{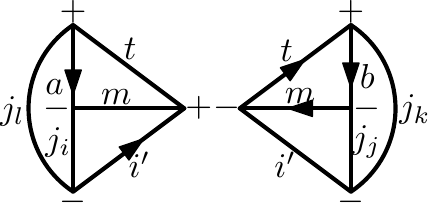}}\notag\\
&=\sqrt{d_t d_{i'}}(-1)^{i'+j_i-j_l}(-1)^{a+m+j_i}(-1)^{i'+j_j+j_k}(-1)^{b+m+j_j}\notag\\
&\qquad\times(-1)^{t+m+i'}
\begin{Bmatrix}
j_i & m & a\\
t & j_l & i'
\end{Bmatrix}
\begin{Bmatrix}
j_j & m & b \\
t & j_k & i'
\end{Bmatrix},
\end{align}
where we used Eq. \eqref{block-3} in the third step, and the identity
\begin{align}
&\makeSymbol{
\includegraphics[height=1.6cm]{figures/4-valent-graph/4-valent-graph-internal-vertex-id-1}}\notag\\
&=(-1)^{i'+j_i-j_l}(-1)^{a+m+j_i}
\makeSymbol{
\includegraphics[height=1.6cm]{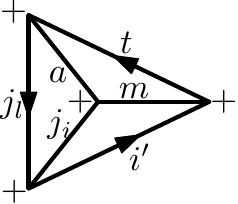}}\notag\\
&\quad\times(-1)^{i'+j_j+j_k}(-1)^{b+m+j_j}(-1)^{t+m+i'}\makeSymbol{
\includegraphics[height=1.6cm]{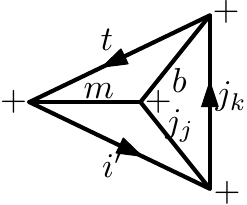}}\notag\\
&=(-1)^{i'+j_i-j_l}(-1)^{a+m+j_i}(-1)^{i'+j_j+j_k}(-1)^{b+m+j_j}(-1)^{t+m+i'}\notag\\
&\qquad\times
\begin{Bmatrix}
j_i & m & a\\
t & j_l & i'
\end{Bmatrix}
\begin{Bmatrix}
j_j & m & b \\
t & j_k & i'
\end{Bmatrix},
\end{align}
in the last step. Thus, the transition amplitude between $\psi_{i'}$ and $\psi_t$ defined in Eq. \eqref{boundary-amplitude} can be calculated as
\begin{align}\label{Z-two-psi}
& \langle\psi_{i'} |{\cal Z}^{\rm EPRL}(\Delta^*)| \psi_t\rangle\notag\\
=&d_{j_i}d_{j_j}d_{j_k}d_{j_l}d_ad_bd_m\makeSymbol{
\includegraphics[height=4.8cm]{figures/4-valent-graph/4-valent-graph-internal-vertex}}\notag\\
&\times\frac{1}{d_{j_i}d_{j_j}d_{j_k}d_{j_l}\sqrt{d_ad_bd_m}}\notag\\
=&\sqrt{d_ad_bd_m} \sqrt{d_t d_{i'}}\notag\\
&\times(-1)^{i'+j_i-j_l}(-1)^{a+m+j_i}(-1)^{i'+j_j+j_k}(-1)^{b+m+j_j}(-1)^{t+m+i'}\notag\\
&\times
\begin{Bmatrix}
j_i & m & a\\
t & j_l & i'
\end{Bmatrix}
\begin{Bmatrix}
j_j & m & b \\
t & j_k & i'
\end{Bmatrix}.
\end{align}

Combining Eqs. \eqref{Z-two-psi} and \eqref{H-matrix-element} yields
\begin{align}\label{eq:4valent-case-cosistency}
&\sum_{\psi_t}\langle\psi_{i'} |{\cal Z}^{\rm EPRL}(\Delta^*)| \psi_t\rangle\langle \psi_t|\hat{H}^{\rm E}_{v_1,e_i,e_j}|\psi_i\rangle\notag\\
=&\sqrt{d_{i'}}\,\chi(1)\chi(m)\chi(j_i)\chi(j_j)\sum_k\sqrt{d_k}\left(\widehat{V^{-1}}_{v_1}\right)_i^k\notag\\
&\times\sum_{c,t}d_t d_c(-1)^{c-i'-m+2t+k-j_i-j_j-j_k-j_l-1}\begin{Bmatrix}
c & m & 1\\
1 & 1 & m
\end{Bmatrix}\notag\\
&\times\sum_a(-1)^{R_a}d_a
\begin{Bmatrix}
c & j_i & a\\
j_i & m & 1
\end{Bmatrix}
\begin{Bmatrix}
j_i & m & a\\
t & j_l & i'
\end{Bmatrix}
\begin{Bmatrix}
t & j_l & a\\
j_i & c & k
\end{Bmatrix}\notag\\
&\times\sum_b(-1)^{R_b}d_b
\begin{Bmatrix}
c & j_j & b\\
j_j & m & 1
\end{Bmatrix}
\begin{Bmatrix}
j_j & m & b \\
t & j_k & i'
\end{Bmatrix}
\begin{Bmatrix}
t & j_k & b\\
j_j & c & k
\end{Bmatrix}\notag\\
=&\sqrt{d_{i'}}\,\chi(1)\chi(m)\chi(j_i)\chi(j_j)(-1)^{i'+m-j_i-j_j-j_k-j_l}\notag\\
&\times\sum_k\sqrt{d_k}\left(\widehat{V^{-1}}_{v_1}\right)_i^k\begin{Bmatrix}
1 & i' & k\\
j_l & j_i & j_i
\end{Bmatrix}\begin{Bmatrix}
1 & i' & k\\
j_k & j_j & j_j
\end{Bmatrix}\notag\\
&\times\sum_t d_t(-1)^t\sum_c(-1)^{R_c}d_c
 \begin{Bmatrix}
1 & m & c\\
t & k & i'
\end{Bmatrix}
\begin{Bmatrix}
t & k & c\\
1 & m & i'
\end{Bmatrix}
\begin{Bmatrix}
1 & m & c\\
m & 1 & 1
\end{Bmatrix}\notag\\
=&\sqrt{d_{i'}}\,\chi(1)\chi(m)\chi(j_i)\chi(j_j)(-1)^{i'+m-j_i-j_j-j_k-j_l}\notag\\
&\times\sum_k\sqrt{d_k}\left(\widehat{V^{-1}}_{v_1}\right)_i^k\begin{Bmatrix}
1 & i' & k\\
j_l & j_i & j_i
\end{Bmatrix}\begin{Bmatrix}
1 & i' & k\\
j_k & j_j & j_j
\end{Bmatrix}\begin{Bmatrix}
i '& i' & 1\\
1 & 1 & k
\end{Bmatrix}\notag\\
&\times\sum_t d_t(-1)^t
\begin{Bmatrix}
i' & i' & 1\\
m & m & t
\end{Bmatrix}\notag\\
=&0,
\end{align}
where in the second and third steps we used \cite{Yutsis:1962bk,Varshalovich:1988ye}
\begin{align}
& \sum_x(-1)^{R_x}d_x
\begin{Bmatrix}
a & b & x\\
c & d & p
\end{Bmatrix}
\begin{Bmatrix}
c & d & x\\
e & f & q
\end{Bmatrix}
\begin{Bmatrix}
e & f & x\\
b & a & r
\end{Bmatrix}\notag\\
=&
\begin{Bmatrix}
p & q & r\\
e & a & d
\end{Bmatrix}
\begin{Bmatrix}
p & q & r\\
f & b & c
\end{Bmatrix}
\end{align}
with
\begin{align}
 R_x:=a+b+c+d+e+f+p+q+r+x,
\end{align}
and in the last step we used \cite{Yutsis:1962bk}
\begin{align}
 \sum_xd_x(-1)^x
\begin{Bmatrix}
j_1 & j_1 & j_3\\
l_1 & l_1 & x
\end{Bmatrix}
=(-1)^{-j_1-l_1}\sqrt{d_{j_1}d_{l_1}}\,\delta_{j_3,0}.
\end{align}
Hence Eq. \eqref{consistency-check} is satisfied for the graph $\gamma$ with a 4-valent vertex. The above calculations can be easily generalized to the case that a graph $\gamma$ has a vertex with valence more that four. The quantum dynamics can be presented by a visual picture as
\begin{align}
 \makeSymbol{
\includegraphics[width=0.45\textwidth]{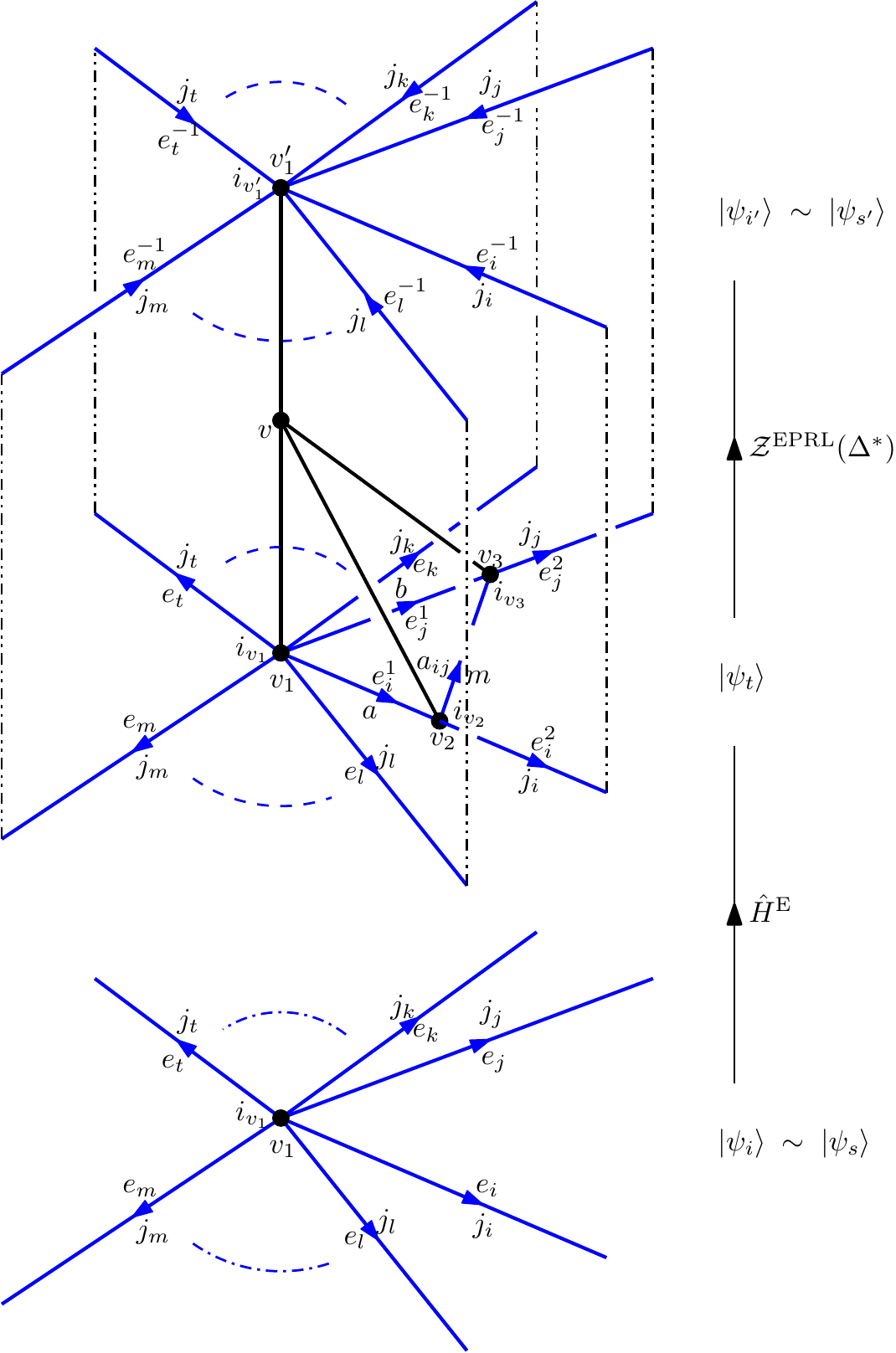}}\notag .
\end{align} 
In this general case, the states appeared in Eq. \eqref{consistency-check} can be graphically expressed by
\begin{align}\label{psi-i-general}
\psi_i&=f_i\makeSymbol{
\includegraphics[height=2cm]{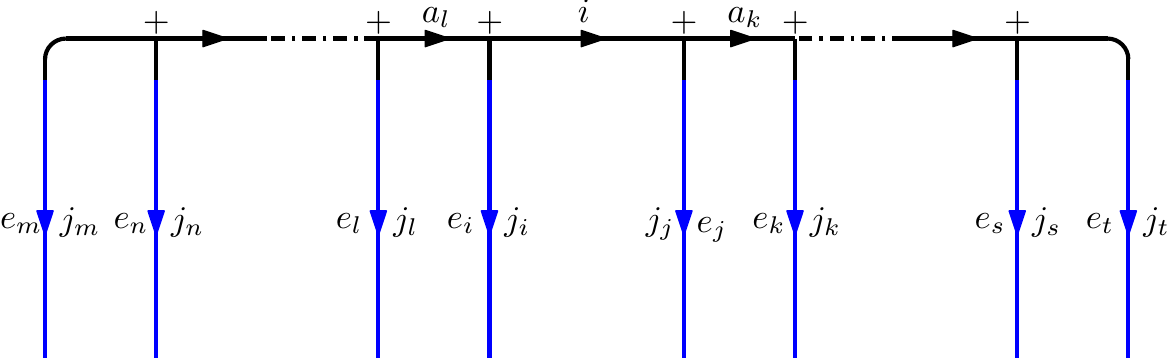}},\\
\psi_t&=f_t\makeSymbol{
\includegraphics[height=2cm]{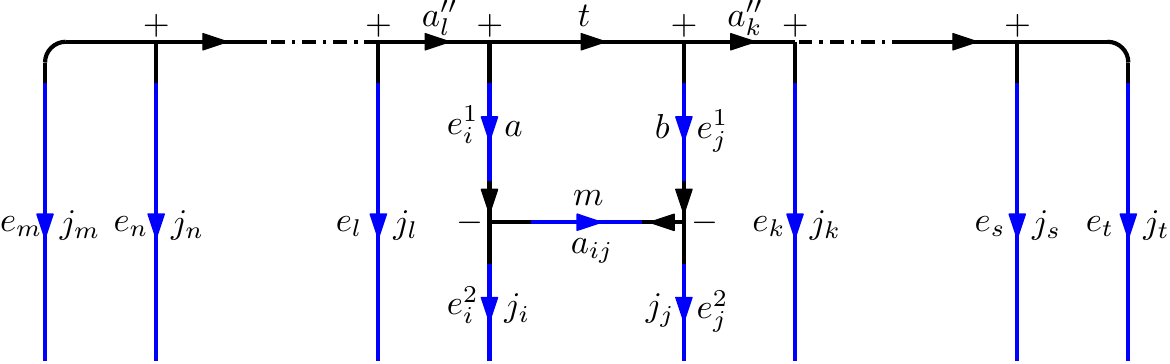}},\label{psi-alpha-general}\\
\psi_{i'}&=f_{i'}\makeSymbol{
\includegraphics[height=2cm]{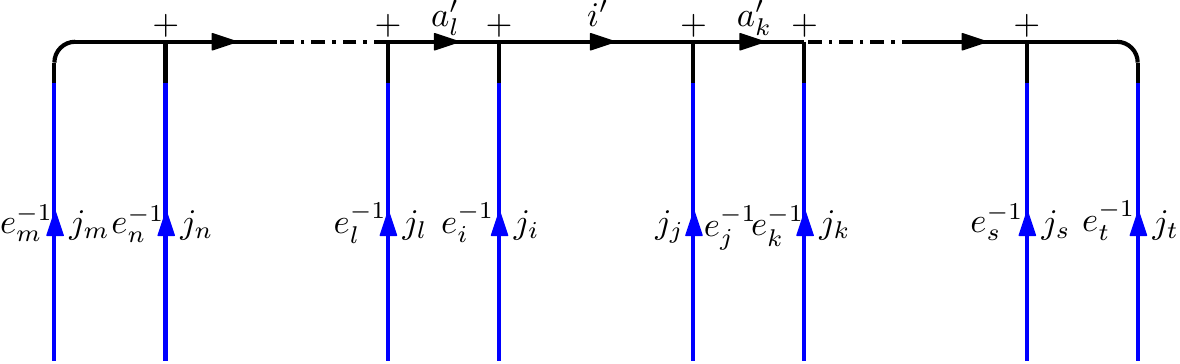}},\label{psi-iprime-general}
\end{align}
where $f_i,f_t,f_{i'}$ denote the normalized factors. Then the transition amplitude between $\psi_{i'}$ and $\psi_t$ defined in Eq. \eqref{boundary-amplitude} can be calculated as
\begin{align}\label{Z-two-psi-general}
& \langle\psi_{i'} |{\cal Z}^{\rm EPRL}(\Delta^*)| \psi_t\rangle\notag\\
=&\sqrt{d_ad_bd_m}\sqrt{\prod_xd_{a_x''}d_t}\sqrt{\prod_yd_{a_y'}d_{i'}}\notag\\
&\times\makeSymbol{
\includegraphics[height=2.6cm]{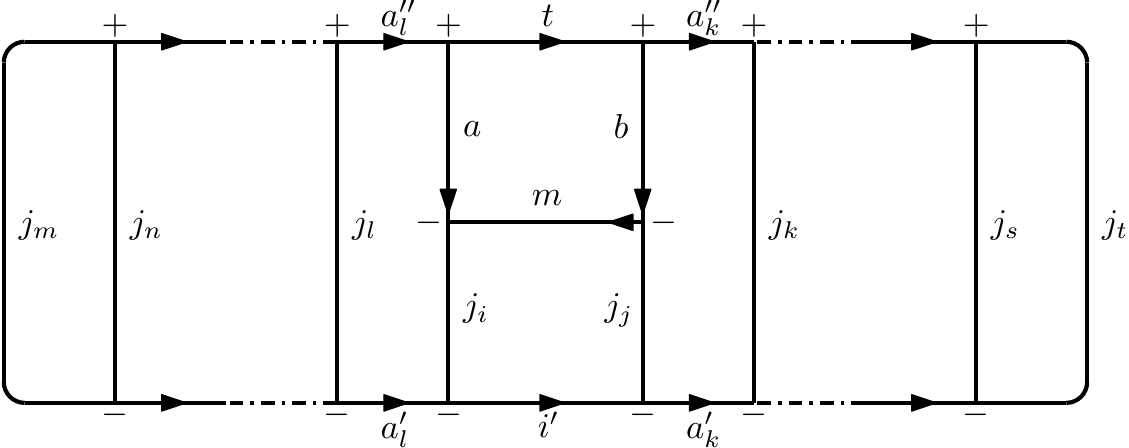}}\notag\\
=&\left(\prod_{x}\delta_{a'_x,a''_x}\right)\sqrt{d_ad_bd_m}\makeSymbol{
\includegraphics[height=4.8cm]{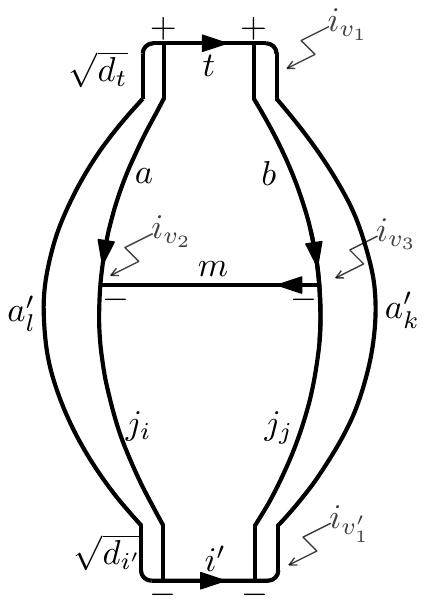}}\notag\\
=&\left(\prod_{x}\delta_{a'_x,a''_x}\right)\sqrt{d_ad_bd_m} \sqrt{d_t d_{i'}}\notag\\
&\times(-1)^{i'+j_i-a'_l}(-1)^{a+m+j_i}(-1)^{i'+j_j+a'_k}(-1)^{b+m+j_j}(-1)^{t+m+i'}\notag\\
&\times
\begin{Bmatrix}
j_i & m & a\\
t & a'_l & i'
\end{Bmatrix}
\begin{Bmatrix}
j_j & m & b \\
t & a'_k & i'
\end{Bmatrix},
\end{align}
where we used Eqs. \eqref{3j-orthogonality-2} and \eqref{two-arrow-cancel} in the second step, and Eq. \eqref{A-v-graph-simplify} in the third step. In this case, the matrix element $\langle\psi_t|\hat{H}^{\rm E}_{v_1,e_i,e_j}|\psi_i\rangle$ reads
\begin{align}\label{H-matrix-element-general}
&\langle\psi_t|\hat{H}^{\rm E}_{v_1,e_i,e_j}|\psi_i\rangle\notag\\
=&\chi(1)\chi(m)\chi(j_i)\chi(j_j)\sum_{\vec{a}'',k}\sqrt{d_k}\left\langle\vec{a}'',k\left|\widehat{V^{-1}}_{v_1}\right|\vec{a},i\right\rangle d_ad_b\notag\\
&\times\sum_cd_c(-1)^{b+c+j_j-1}
\begin{Bmatrix}
c & j_i & a\\
j_i & m & 1
\end{Bmatrix}
\begin{Bmatrix}
c & j_j & b\\
j_j & m & 1
\end{Bmatrix}
\begin{Bmatrix}
c & m & 1\\
1 & 1 & m
\end{Bmatrix} 
 \notag\\
&\times d_t(-1)^{a''_l-j_i-k}(-1)^{t+b-a''_k}
\begin{Bmatrix}
t & a''_l & a\\
j_i & c & k
\end{Bmatrix}
\begin{Bmatrix}
t & a''_k & b\\
j_j & c & k
\end{Bmatrix}\notag\\
&\times\frac{1}{\sqrt{d_ad_bd_md_t}},
\end{align}
where $|\vec{a},i\rangle$ denotes the normalized intertwiner of $\psi_i$ in Eq. \eqref{psi-i-general}, $\vec{a}$ denotes the set of intermediate coupling spins other than $i$, and $|\vec{a}'',k\rangle$ is the one obtained from $|\vec{a},i\rangle$ by replacing $\vec{a}$ by $\vec{a}''$ and $i$ by $k$. Combining Eqs. \eqref{Z-two-psi-general} and \eqref{H-matrix-element-general}, we have
\begin{align}\label{eq:general-case-cosistency}
&\sum_{\psi_t}\langle\psi_{i'} |{\cal Z}^{\rm EPRL}(\Delta^*)| \psi_t\rangle\langle \psi_t|\hat{H}^{\rm E}_{v_1,e_i,e_j}|\psi_i\rangle\notag\\
=&\sqrt{d_{i'}}\,\chi(1)\chi(m)\chi(j_i)\chi(j_j)\sum_k\sqrt{d_k}\left\langle\vec{a}',k\left|\widehat{V^{-1}}_{v_1}\right|\vec{a},i\right\rangle\notag\\
 &\times\sum_{c,t}d_t d_c(-1)^{c-i'-m+2t+k-j_i-j_j-a'_k-a'_l-1}\begin{Bmatrix}
c & m & 1\\
1 & 1 & m
\end{Bmatrix}\notag\\
&\times\sum_a(-1)^{R_a}d_a
 \begin{Bmatrix}
c & j_i & a\\
j_i & m & 1
\end{Bmatrix}
\begin{Bmatrix}
j_i & m & a\\
t & a'_l & i'
\end{Bmatrix}
\begin{Bmatrix}
t & a'_l & a\\
j_i & c & k
\end{Bmatrix}\notag\\
&\times\sum_b(-1)^{R_b}d_b
\begin{Bmatrix}
c & j_j & b\\
j_j & m & 1
\end{Bmatrix}
\begin{Bmatrix}
j_j & m & b \\
t & a'_k & i'
\end{Bmatrix}
\begin{Bmatrix}
t & a'_k & b\\
j_j & c & k
\end{Bmatrix}\notag\\
=&0.
\end{align}
Therefore, in the general case, the quantum dynamics between the covariant and canonical LQG, determined by the generalized Euclidean EPRL model and by the Hamiltonian constraint operator $\hat{H}^{\rm E}_\delta(N)$ respectively, are consistent to each other on the spin network states with one vertex in the sense of Eq. \eqref{quantum-dynamical-relation} for $\beta=1$. Although the above discussion is confined to the case that the interior vertex $v\in\Delta^*$ is only 4-valent, and thus is not dual to a 2-cell complex $\Delta$, it is straightforward to extend it to the case of higher valent internal vertex which does have a geometric interpretation of certain 2-cell $\Delta^*$. For instance, a higher valent interior vertex $v\in\Delta^*$ can be obtained by adding some vertices (and edges) to the boundary graph of $\Delta^*$ associated to $\psi_{i'}$ as
\begin{align}
 \makeSymbol{
\includegraphics[width=0.45\textwidth]{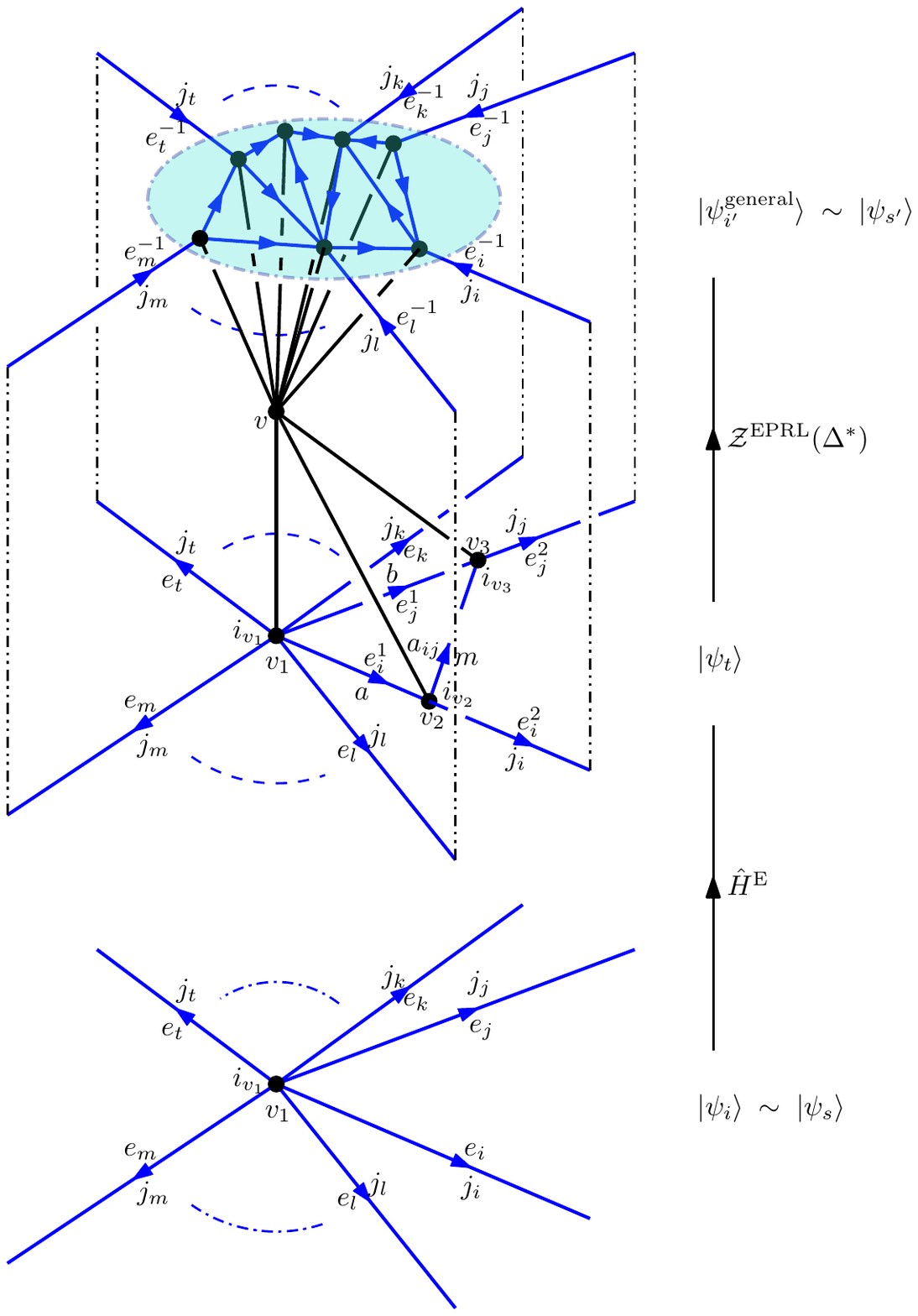}}\notag .
\end{align}
Interestingly, the consistency of the quantum dynamics still holds for the above generalized $\psi_{i'}^{\rm general}$. The key observation comes from the fact that Eq. \eqref{eq:general-case-cosistency} holds for the vertex amplitude of $\Delta^*$ since it contains the same quantities as Eq. \eqref{Z-two-psi-general} determined only by the local information, i.e., the intertwiners associated to $v_1$ and to the new created vertices $v_2$ and $v_3$ by the action of $\hat{H}^{\rm E}$ at $v_1$, while the rest does not depend on the spins $a,b$ and $t$. To see this, let us calculate the transition amplitude for the generalized $\psi_{i'}^{\rm general}$, which is given by
\begin{align}\label{Z-two-psi-general-new}
& \langle\psi_{i'}^{\rm general} |{\cal Z}^{\rm EPRL}(\Delta^*)| \psi_t\rangle\notag\\
=&\sqrt{d_ad_bd_m}\sqrt{\prod_xd_{a_x''}d_t}\notag\\
&\times\makeSymbol{
\includegraphics[width=7.2cm]{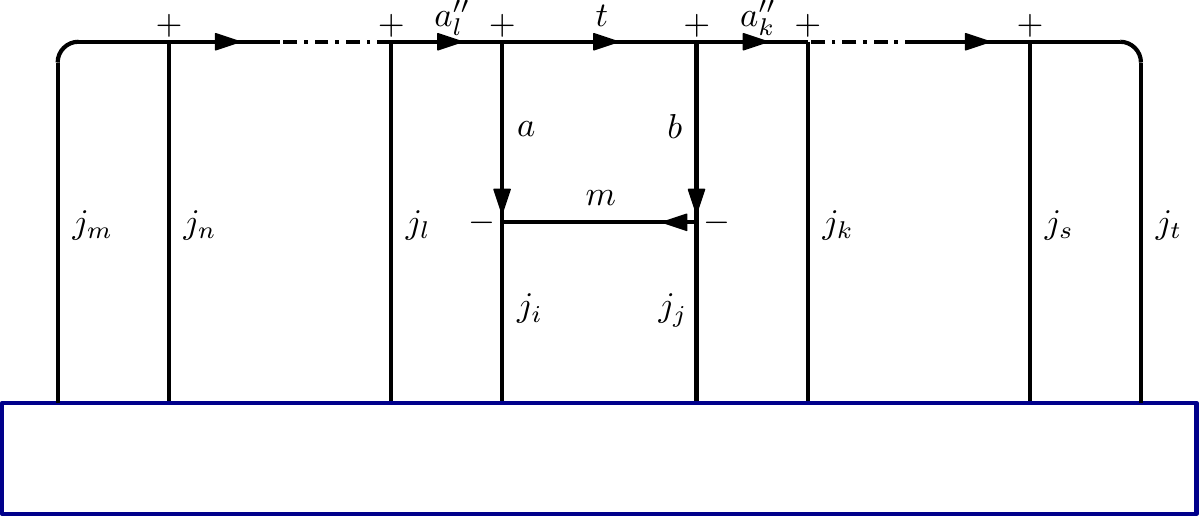}}\notag\\
=&\sum_{\cdots,a'_l,i',a'_k\cdots}\sqrt{d_ad_bd_m}\sqrt{\prod_xd_{a_x''}d_t}\prod_yd_{a_y'}d_{i'}\notag\\
&\times\makeSymbol{
\includegraphics[width=7.2cm]{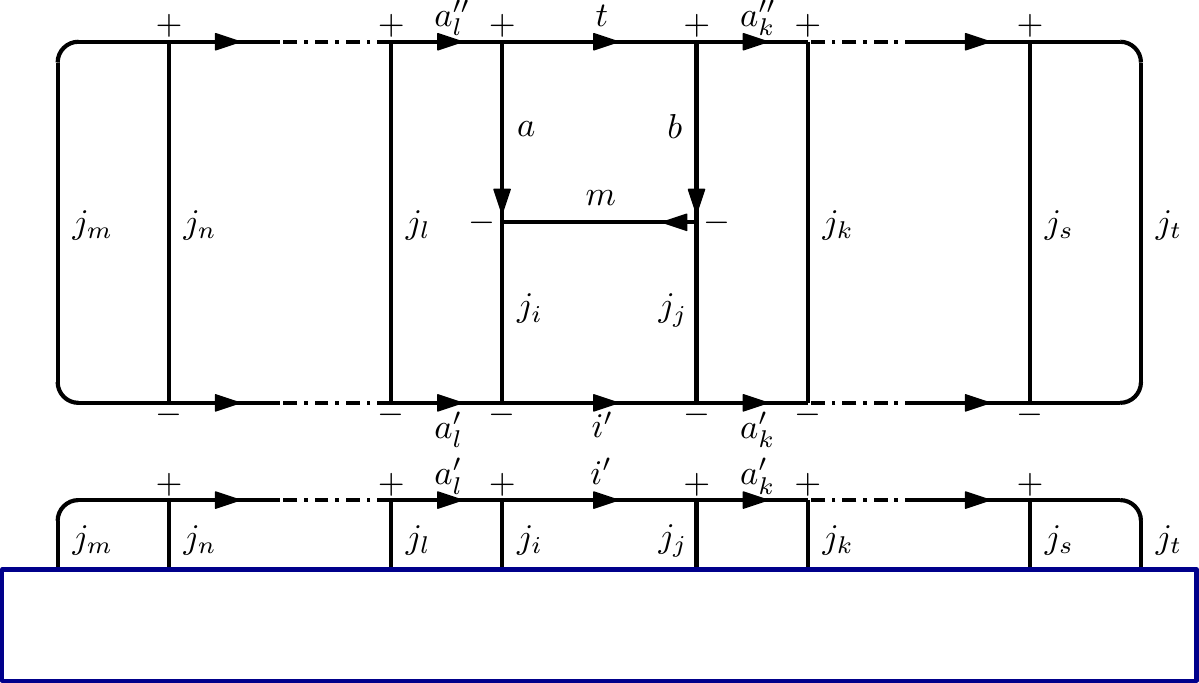}}\notag\\
=&\sum_{\cdots,a'_l,i',a'_k\cdots}\sqrt{\prod_yd_{a_y'}d_{i'}} f(a'_l,i',a'_k,\cdots)\langle\psi_{i'}|{\cal Z}^{\rm EPRL}(\Delta^*)| \psi_t\rangle,
\end{align}
where the block diagram with external lines represents the contraction of intertwiners in $\psi_{i'}^{\rm general}$ associated to vertices related to the interior vertex $v\in\Delta^*$, in the second step the graphical rule \eqref{block-4} was used, in the third step Eq. \eqref{Z-two-psi-general} was used, and $f(a'_l,i',a'_k,\cdots)$ denotes a factor corresponding to the block diagram in second equality, which does not depend on the spins $a$, $b$, and $t$. As a direct consequence, Eq. \eqref{Z-two-psi-general-new} is proportional to \eqref{Z-two-psi-general} with a factor independent of the spins $a,b$ and $t$. This leads to
\begin{align}\label{eq:general-case-cosistency-new}
 \sum_{\psi_t}\langle\psi_{i'}^{\rm general} |{\cal Z}^{\rm EPRL}(\Delta^*)| \psi_t\rangle\langle \psi_t|\hat{H}^{\rm E}_{v_1,e_i,e_j}|\psi_i\rangle=0.
\end{align}
It is easy to see that our calculation indicates the consistency for the general dual 2-cell complex $\Delta^*$ between $\psi_t$ and $\psi^{\rm general}_{i'}$  (or $\psi_{s'}$) provided that each triple of vertices (e.g., $v_1$, $v_2$ and $v_3$) associated to $\psi_t$, belonging to a loop $\alpha_{ij}$ produced by $\hat{H}^{\rm E}_{v_1,e_i,e_j}$, is related to an internal vertex (e.g., $v$) by internal edges of $\Delta^*$. Thus the consistency \eqref{quantum-dynamical-relation} is valid generally, and it indicates that the EPRL model can provide a rigging map \eqref{rigging-map} for the Hamiltonian constraint operator $\hat{H}^{\rm E}_\delta(N)$ in Eq. \eqref{eq:canonical-dynamics}.

\section{Summary and discussion}
\label{sec-V}
A major challenge in LQG is how to relate its covariant formulation to its canonical formulation in quantum dynamics. In previous sections, we studied the relation by taking the viewpoint that SFM  provides a rigging map such that the Hamiltonian constraint in canonical LQG is weakly satisfied. This idea was first proposed in \cite{Alesci:2011ia}, where the consistency between the EPRL SFM and the Euclidean Hamiltonian constraint operator proposed by Thiemann in \cite{Thiemann:1996aw} was checked. While the same EPRL SFM is concerned here, the Hamiltonian constraint operator which we considered is the Euclidean version $\hat{H}^{\rm E}(N)$ proposed in \cite{Yang:2015zda}. The virtue of $\hat{H}^{\rm E}(N)$ is that it is well defined in certain partially diffeomorphism invariant Hilbert space and can be promoted to a symmetric operator. 

The graphical calculus was used as a powerful tool to give direct and concise derivations of the partition function $ {\cal Z}^{\rm EPRL}(\Delta^*)$ in Eqs. \eqref{resulting-Z-nonboundary} [or \eqref{resulting-Z-nonboundary-algebraic}] and \eqref{resulting-Z-boundary} [or \eqref{resulting-Z-boundary-algebraic}] of the generalized Euclidean EPRL model for $\Delta^*$ without and with a boundary respectively, as well as the matrix elements \eqref{H-matrix-element} [or Eq. \eqref{H-matrix-element-general}] of the Hamiltonian constraint operator $\hat{H}^{\rm E}_\delta(N)$ on certain spin network states. Our result of Eq. \eqref{eq:general-case-cosistency-new} shows that in the Euclidean case the generalized EPRL model can provide a rigging map such that the Hamiltonian constraint operator proposed in \cite{Yang:2015zda} is weakly satisfied on the spin network states with one vertex for the Immirzi parameter $\beta=1$. Hence, in this sense, the quantum dynamics between covariant LQG and canonical LQG are consistent to each other for these states. Moreover, we showed how to generalize the graphical calculus to the calculations of SFMs. It provides a visual and powerful tool alternative to the algebraic one. 

It should be noted that the Hamiltonian constraint operator $\hat{H}^{\rm E}(N)$ is well defined in the Hilbert space ${\cal H}_{\rm np4}$ consisting of the almost diffeomorphism invariant states obtained by group-averaging the diffeomorphisms of $\Sigma$ but leaving fixed sets of nonplanar vertices with valence higher than three. Though the results of Eqs. \eqref{eq:4valent-case-cosistency}, \eqref{eq:general-case-cosistency}, and \eqref{eq:general-case-cosistency-new} were derived with the kinematical Hilbert space ${\cal H}_{\rm kin}$, there is no obstacle to promote them to ${\cal H}_{\rm np4}$ since the partially diffeomorphism transformations neither change the relevant vertices nor change the intertwiners and spins on the graphs. Hence, our results indicate actually the consistency between the partially diffeomorphism invariant generalized EPRL SFM and $\hat{H}^{\rm E}(N)$ in ${\cal H}_{\rm np4}$ of canonical LQG. The fact that the rigging map given by the EPRL SFM can weakly satisfy both Thiemann's Hamiltonian  $\hat{H}^{\rm E}_{\rm T}(N)$ in \cite{Thiemann:1996aw} and the Hamiltonian constraint $\hat{H}^{\rm E}(N)$ in \cite{Yang:2015zda} manifests that the physical states provided by the rigging map on the spin network states with one vertex does not include all the solutions to  $\hat{H}^{\rm E}_{\rm T}(N)$ or $\hat{H}^{\rm E}(N)$. Further investigation is still desirable to reveal more accurate relations between the covariant and canonical dynamics of LQG. Note also that our discussion is confined to the case of $\beta=1$, it is desirable to generalize the calculations to the general cases of $\beta\neq 1$ and even to the Lorentzian signature. Although the generalization would be nontrivial, it could be handled in principle. We leave these issues for future study.

\begin{acknowledgments}
This work is supported in part by NSFC Grants No. 11765006, No. 11961131013 and No. 11875006. C. Z. acknowledges the support by the Polish Narodowe Centrum Nauki, Grant No. 2018/30/Q/ST2/00811.
\end{acknowledgments}


%

\end{document}